\documentclass[preprint,12pt,authoryear]{elsarticle}
\usepackage[letterpaper,top=2cm,bottom=2cm,left=3cm,right=3cm,marginparwidth=1.75cm]{geometry}
\usepackage{amssymb}
\usepackage{amsmath}
\usepackage{graphicx}
\usepackage{dirtytalk}
\usepackage[colorlinks=true,linkcolor=blue]{hyperref}
\usepackage{multicol}
\usepackage{comment}
\usepackage{caption}
\usepackage{subcaption}
\usepackage{algorithm}
\usepackage{algpseudocode}
\usepackage{physics}
\usepackage[autostyle]{csquotes}
\usepackage{dirtytalk}
\usepackage{nicefrac}
\usepackage{overpic}
\date{}
\usepackage{lineno}
\newcommand{\mbs}[1]{\boldsymbol{#1}}

\def\bA{{\mbs{A}}}  \def\bC{{\mbs{C}}}
 \def\bE{{\mbs{E}}} \def\bF{{\mbs{F}}}
\def\bG{{\mbs{G}}}  \def\bI{{\mbs{I}}}
  
 \def\bN{{\mbs{N}}} 
\def\bP{{\mbs{P}}}  \def\bR{{\mbs{R}}}
  
 \def\bX{{\mbs{X}}} \def\bY{{\mbs{Y}}}
 \def\b0{{\mbs{0}}}
\def\ba{{\mbs{a}}}  
 \def\be{{\mbs{e}}}

 \def\bn{{\mbs{n}}} 
  
  \def\bu{{\mbs{u}}}
 \def\bx{{\mbs{x}}} \def\by{{\mbs{y}}}

\def    \tr {\textrm{tr\,}}

\def    \uth {{u_s^{\rm th}}}

\begin{document}

\begin{frontmatter}

\title{Intestinal peristalsis and wrinkling: A novel paradigm}

\author[inst1,inst2]{René Thierry Djoumessi}
\author[inst1]{Christopher Miller}
\author[inst3]{Nipuni D. Nagahawatte}
\author[inst2]{Marco Paggi}
\author[inst3]{Leo K. Cheng}
\author[inst1]{Alessio Gizzi\corref{cor1}}

\affiliation[inst1]{organization={Department of Engineering, Università Campus Bio-Medico di Roma},
            addressline={Via A. del Portillo 21}, 
            city={Rome},
            postcode={00128}, 
            country={Italy}}

\affiliation[inst2]{organization={IMT School for Advanced Studies Lucca},
            addressline={Piazza San Francesco 19}, 
            city={Lucca},
            postcode={55100}, 
            country={Italy}}

\affiliation[inst3]{organization={Auckland Bioengineering Institute, University of Auckland},
            addressline={ Level, 6/70 Symonds Street, Grafton}, 
            city={Auckland},
            postcode={1010}, 
            country={New Zealand}}    

\cortext[cor1]{Corresponding author Email addresses: a.gizzi@unicampus.it (A. Gizzi)}

\begin{abstract}
A new computational framework for modeling the intestinal wall as a multi-layered fiber–reinforced continuum is presented. The framework reproduces for the first time physiological motility and overcoming large-displacements limitations (self-contact and volume locking) occurring in classical hyperelastic formulations of soft tissues. We introduce: i) layer-specific functions, segregating active circumferential and longitudinal muscle fibers while maintaining homogeneous passive reinforcement, and ii) a quasi-incompressible volumetric contribution, to handle large peristaltic contractions. Cell electrophysiology is further extended to reproduce both slow waves and spike bursting activities thus mimicking for the first time a localized neural excitation in a three-dimensional geometry of small intestine segment. We introduce a spatio-temporal modulation of contractility to accurately capture activation driven by both slow waves and spike bursts. The overall coupled nonlinear electromechanical boundary valued problem is modeled following the active strain approach. A robust augmented-Lagrangian contact algorithm is also embedded to avoid self-penetration and geometrical instabilities under large displacements. The 8-variables nonlinear governing equations are then discretized using in house $\mathcal{P}_1$-$\mathcal{P}_2$-$\mathcal{P}_4$ finite elements codes implemented within the \texttt{GetFEM} library. Numerical experiments demonstrate the ability of the proposed framework to reproduce physiological peristalsis, i.e., wall contraction $>80\%$, thus allowing full occlusion matching {\it in vivo} endoscopic images, and naturally generating wrinkling patterns consistent with experimental observations. We show that an active electromechanics anisotropic heterogeneous modeling strategy is critical for a numerically stable and physiologically accurate representation of gastrointestinal motility.
\end{abstract}


    \end{frontmatter}

\section{Introduction}

The gastrointestinal tract is a highly dynamic system of organs whose function emerges from the close interplay of mechanical, biochemical, and electrophysiological processes \citep{azzouz2018physiology, sulaiman2019mri, precup2019gut}. Beyond the transport of luminal contents, the intestine mixes, breaks down, and conditions chyme to optimize nutrient absorption. This functionality is reliant on a wide repertoire of coordinated motions, including peristalsis, segmentation, and strong occlusive contractions, that together give rise to local pressure variations, shear zones, and large wall deformations \citep{kuruppu2024electromechanical}. Additionally, the highly structured luminal morphology, characterized by folds and villi, further enhances nutrient absorption and modulates local flow patterns \citep{zha2021role}. These phenomena are not passive: they arise from an intrinsically electromechanical system in which pacemaker activity from interstitial cells of Cajal (ICC), coupled with smooth muscle cells (SMC), drives gut motility and induces spatially heterogeneous mechanical states.

\begin{figure}[h]
%\centering\includegraphics[width=\textwidth]{images/fiberkim1.pdf}
\centering\includegraphics[width=140mm]{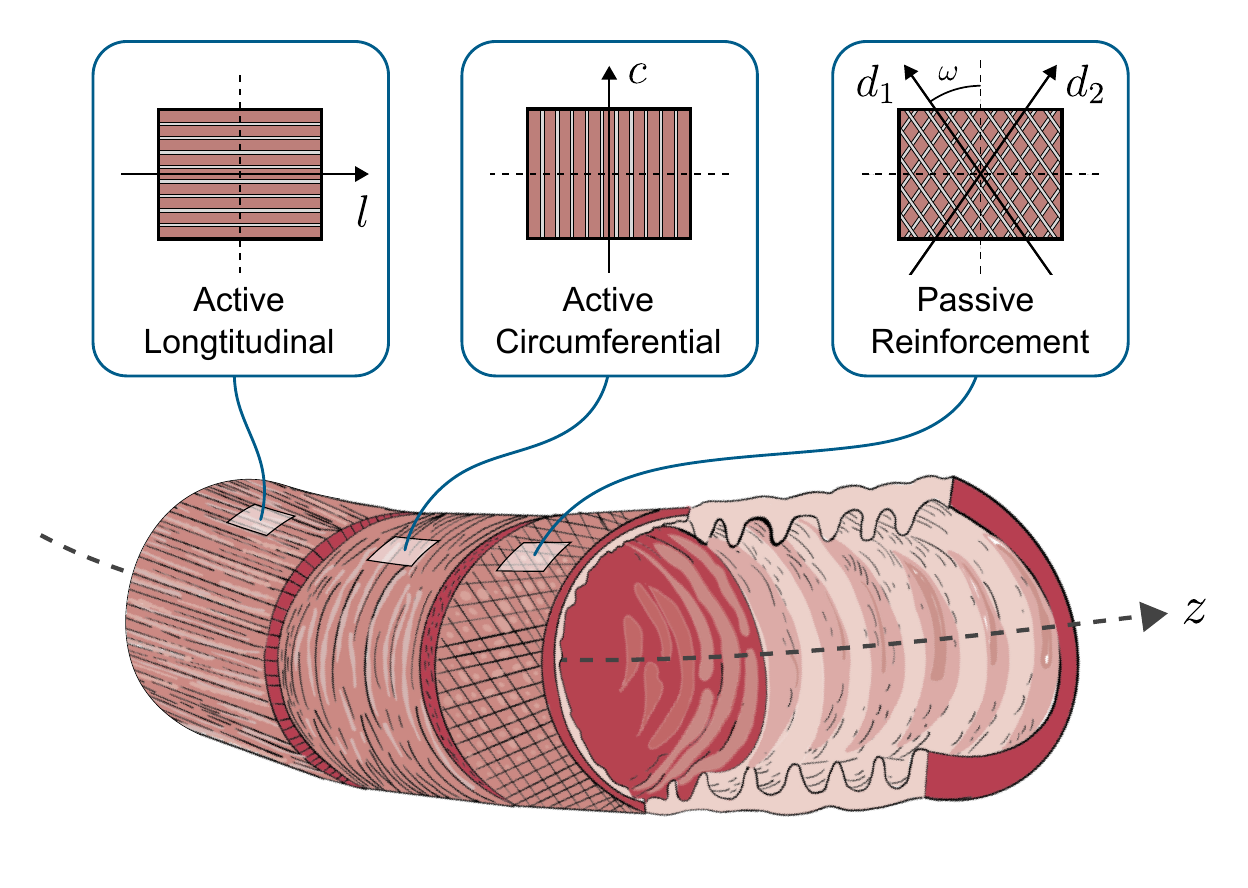}
    \caption{Schematic diagram of the fibers within the intestinal wall, demonstrating how the GI wall is composed of four families of fibers embedded in an otherwise isotropic elastin matrix.  $l$ represents the external longitudinal muscular layer, $c$ is the internal circumferential muscular fiber, and $d_1,\,d_2$ are additional reinforcement defined with respect to the circumferential direction by the angle $\omega$.}
    \label{generirfiber}
\end{figure}

The intestinal wall exhibits a complex multilayered architecture, consisting of the mucosa, submucosa, and the mechanically significant circular and longitudinal muscle layers, each characterized by distinct anisotropic fiber orientations \citep{sokolis2021variation}, as schematically presented in Fig.~\ref{generirfiber}. Circular fibers are fundamental to the radial occlusion of the gastrointestinal wall, whilst longitudinal fibers contribute to the axial propulsion of intraluminal contents. Furthermore, the presence of passive reinforcing fibers enables effective interlaminar load transmission. Electrophysiologically, slow waves generated by ICC propagate along the gut and set the basal active rhythm; this may produce high-amplitude contractions upon appropriate stimuli \citep{cheng2013mapping}. Intense occlusive motility typically arises from excitatory inputs delivered by the enteric nervous system (ENS) through spike bursts, which may be myogenic (slow‑wave–associated) or neurogenic (slow‑wave–independent) \citep{fung2020functional, kuruppu2021high}. The multiscale interaction between intestinal microstructure, spatiotemporal electrophysiology, and soft tissue biomechanics ultimately governs the ability of the gut to generate peristalsis, achieve luminal occlusion, and develop surface instabilities such as wrinkling \citep{pathak2019double}.

Surface instabilities have traditionally been studied using simplified film–substrate configurations \citep{nguyen2020wrinkling}, emphasizing the role of contrasting stiffness, growth mismatch, and adhesion \citep{wang2015three}. Whilst these models rely on externally imposed compression, they highlight stiffness heterogeneity as a key driver \citep{stewart2016wrinkling}. Under certain conditions, the intestinal wall alternates between peristaltic and wrinkling regimes, both of which represent physiological states. This observed phenomenology suggests an underlying mechanical instability, although its precise characterization in soft biological tissues remains unclear.

Significant progress has been made in the electromechanical constitutive modeling of the gastrointestinal wall \citep{patel2022biomechanical}. Recently, we considered four distinct fiber families (two active and two passive), thus providing an advanced description of intestinal electromechanics \citep{DJOUMESSI2024116989}. However, the approach had two key limitations: (1) slow wave and spike-burst activities were not distinguished; and (2) the material model was homogeneous, with all fiber families assumed to be present throughout the wall thickness, leading to an artificial extension of longitudinal fiber contractility into the circular layer, and vice versa. Such a modeling approach results in non-physiological ``mechanical competition" between wall layers, limiting luminal occlusion to approximately $20$–$30\%$, thus preventing the effective characterization of wrinkling and luminal self-contact. The formulation was further extended in \citep{djoumessi2026self} to account for external self-contact.

In the present work, the limitations above were addressed, and we generalize the electromechanical framework by incorporating: (i) a multilayered architecture with layer-specific fiber families; (ii) the coexistence of slow waves and spike bursts; and (iii) internal (luminal) self-contact, a crucial mechanism for proper physiological intestinal function. For the first time, we demonstrate that the interplay between slow waves, spike bursts, and anisotropic wall heterogeneity can induce either peristaltic contractions or wall wrinkling even in the absence of external loading. These findings pave the way towards a new paradigm in soft active media, in which wrinkling instabilities are not merely passive mechanical artifacts but rather functional phenomena driven by the underlying active microstructure. 

The manuscript is organized as follows: Sec.~\ref{sec:model} presents the electromechanical model; Sec.~\ref{application} applies the model to the simulation of intestinal motility and compares the results with experimental observations; Sec.~\ref{sec:optimal_control} identifies material parameters that optimally reproduce physiological contraction patterns; Sec. \ref{quantitative_r} gives a quantitative assessment of the luminal occlusion; and Sec.~\ref{conclu} summarizes the main findings and discusses the models limitations.

%%%%%%

\section{Intestine electromechanics and contact}
\label{sec:model}
In this section, we briefly recall the governing equations for the coupling of GI electrophysiology and active-strain finite deformation. We also introduce the methods used to incorporate spike bursts and layer-specific stiffness within the constitutive modeling framework.

Scalars,  vectors, and second-order tensors are represented with lowercase letters ($a$), lowercase bold letters ($\ba$), and capital bold letters ($\bA$), respectively, where ($\bA^T$) denotes the transpose of a tensor. Following standard tensor notation, we indicate the scalar product with $(\cdot)$, the double contraction with $(:)$, and the dyadic product with $(\otimes)$. Moreover, $\nabla$, $\nabla\cdot$, and $\nabla^2$ represent the gradient, divergence, and Laplace operator, respectively.

\subsection{Kinematics}
GI tissue is assumed to behave as an active deformable solid, for which a finite elasticity framework of nonlinear continuum mechanics holds \citep{holzapfel2002nonlinear}. Let $\bX$ be the position vector in the undeformed (material) configuration $\Omega_0$, and $\bx$ be the position vector in the deformed (current) configuration $\Omega$. The deformation gradient tensor and its associated Jacobian are therefore defined as $\bF={\partial\bx}/{\partial\bX}$ and $J=\det\bF > 0$, respectively, with $\bC=\bF^T \bF$ denoting the right Cauchy-Green deformation tensor. The first isotropic invariant of the deformation is $I_1(\bC)=\tr(\bC)$, where $\tr( \cdot )$ is the trace operator, and the fourth anisotropic pseudo-invariant is $I_4(\bC)=\bC:\bG$, where $\bG$ denotes a structural tensor defining fiber direction.

Intestinal contraction arises from the strong nonlinear coupling between the tissues' electrophysiological cellular spatiotemporal dynamics and its hyperelastic mechanical response. The active strain approach has been shown to provide a robust and effective framework for describing this relationship. Specifically, the deformation gradient tensor is multiplicatively decomposed into an elastic, $\bF_e$, and an inelastic part, $\bF_a$, such that the total deformation gradient reads as $ \bF = \bF_e \bF_a \,$. Accordingly,  we introduce an exact decomposition of the elastic gradient tensor into volumetric and isochoric parts: 
\begin{equation} \label{eq:1}
    \bF = \bF_e \bF_a, \quad \bar{\bF_e} = J_e^{-1/3}\bF_e \,.
\end{equation}
We consider longitudinal and circumferential SMC directions as contractile units governed by:
\begin{equation} \label{eq:2}
    \bF_a = \bI-\gamma(u_s)(\alpha_c \bn_c \otimes  \bn_c + \alpha_l \bn_l \otimes \bn_l ) + \gamma_n \bn_n \otimes \bn_n \,,
\end{equation}
where $\bn_c$ and $\bn_l$ are the orthonormal unit vectors defining the circumferential and longitudinal directions in the reference configuration, respectively \citep{DJOUMESSI2024116989, djoumessi2026self}. The relation $\bn_n = \bn_c \times \bn_l$ defines the unit vector orthogonal to their plane, whilst $\alpha_c$ and $\alpha_l$ are material parameters (or contractility) governing the degree of circumferential and longitudinal active shortening, respectively. The term $\gamma_n$ derives from the condition $\det (\bF_a) =1$, and as such, reads:
\begin{equation} \label{eq:3}
   \gamma_n = 
   \frac{1-\left(1-\gamma(u_s)\alpha_c\right)\left(1-\gamma(u_s)\alpha_l\right)}{\left(1-\gamma(u_s)\alpha_c\right)\left(1-\gamma(u_s)\alpha_l\right)} \,.
\end{equation}
The excitation function $\gamma(u_s)$ couples the mechanical problem with the electrophysiological one via a smooth activation function, dependent on the SMC transmembrane potential $u_s$ (see the next subsection): 
\begin{equation}\label{gammafun}
    \gamma(u_s) = (1-e^{-\beta_1(u_s-V_{th})})(1-e^{-\beta_2(u_s-u_s^{th})}) H(u_s-\uth),
\end{equation}
where $\beta_1$, $\beta_2$, and $u_s^{th}$ are the material parameters linked to SMC intracellular calcium (Ca$^{2+}$) dynamics, and $H(u_s-\uth)$ is a Heaviside step function that activates the contraction whenever the threshold $\uth$ is reached. To ensure that ﬁbers cannot contract to zero or negative lengths, we enforce that each factor in the denominator of Eq.~\eqref{eq:3} is positive \citep{brandstaeter2018computational}. 

\begin{figure}[h]
\centering\includegraphics[width=\textwidth]{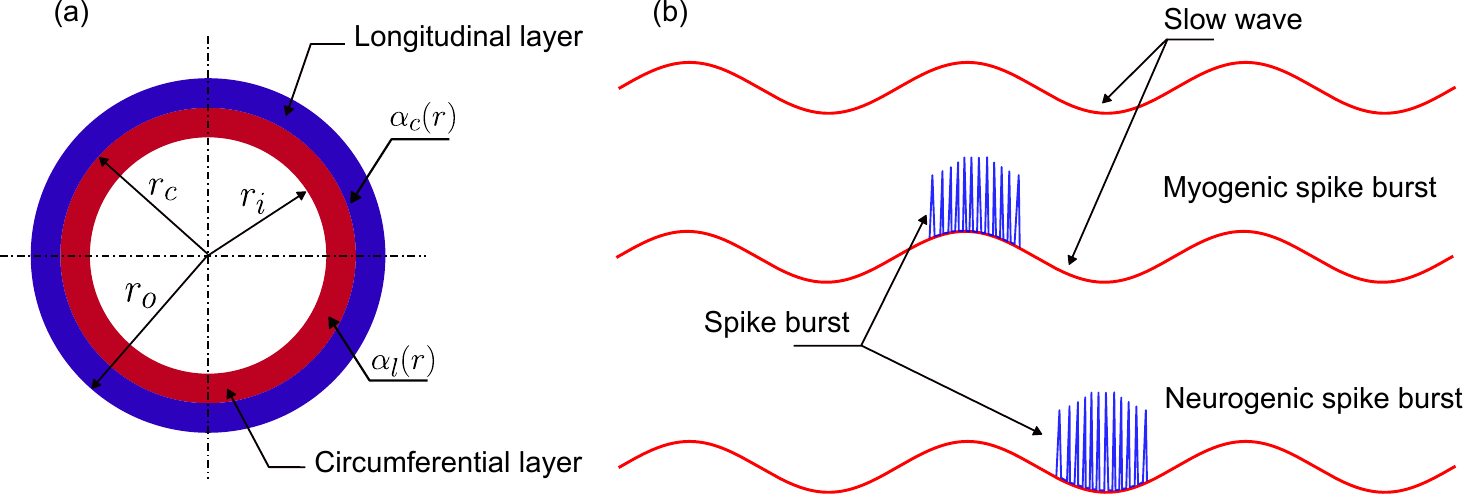}
    \caption{Schematic representation of the
(a) layer-specific distinction for material parameters;
(b) difference between spike bursts and slow waves.}
    \label{generirepr}
\end{figure}

Crucially, in accordance with widespread histological observations of the intestinal wall, smooth muscle fiber activation is prescribed in a layer-specific manner (see Fig.~\ref{generirepr}a) through the contractility parameters $\alpha_l(r)$ and $\alpha_c(r)$ (detailed subsequently), which are assumed to be a function of the radial distance through the GI wall, $r$. The numerical values for model parameters used in this work are provided in Tab.~\ref{tab:active}.

%\paragraph{Remark} Recent experimental observations highlight the influence of stimulus duration on intestinal wall motility, demonstrating that longer stimulation leads to stronger contractions \citep{nagahawatte2023evaluation}. Accordingly, we introduce space- and time-dependent contractility functions, referred to as \emph{effective contractilities}, as detailed in Sec.~\ref{contract_model}.

\subsection{Electrophysiology}
Here, we provide a brief overview of the mathematical representation of slow waves, followed by our novel approach to the incorporation of spike bursts within the existing framework

\subsubsection{Slow Waves}
\label{sec:slowW}
%Intestinal motility is governed by slow wave activity arising from the interaction between ICCs and SMCs.
%\citep{huizinga2009gut,sanders2016regulation,sanders2006interstitial}. 
Signals generated by ICCs propagate through gap junctions \citep{hanani2005intercellular}, inducing calcium influx in SMCs via voltage-gated channels \citep{sanders2016regulation, corrias2008quantitative}. This process ultimately triggers smooth muscle contraction and the resulting baseline motility of the intestinal wall. Several mathematical models have been developed to reproduce the fundamental characteristics of GI slow waves~\citep{Cheng2010}. In this study, we adopt a four-variable phenomenological model of intestinal electrophysiology \citep{aliev2000simple,gizzi2010electrical}:
{
\begin{align}
    \frac{\partial u_q}{\partial t} 
    &= 
    f(u_q)
    +
    D_q \nabla^2 u_q
    -
    v_q + F_q(u_q) \quad 
    %\textrm{on} \quad \Omega_0 \times [0,T], 
    \\
    \frac{\partial v_q}{\partial t}
    &= 
    \epsilon_q[\lambda_q (u_q-\beta_q)-v_q] 
    \quad \textrm{on} \quad \Omega_0 \times [0,T],
    \quad \textrm{for}
    \quad q=s,i \,,
\end{align}
}
consisting of two coupled reaction-diffusion systems that ensure rhythmic oscillations in time and nonlinear wave propagation in space, where the indices $q=s,i$ represent SMC's and ICC's, respectively (see \ref{electrophy_mo} for further details). Critically, this approach captures the key entrainment properties associated with GI tissue's excitation-contraction mechanism, as discussed in \citep{brandstaeter2018computational}.

%The electrophysiological component is used to impose a peristaltic activation pattern. Then according to our electrophysiology model, the Eq.~\eqref{gammafun} becomes: 
%\begin{equation}\label{gammafun1}
%    \gamma(u_s) = (1-e^{1-\beta_1(u_s-V_{th})})(1-e^{1-\beta_2(u_s-V_{th})}) H(u_s-V_{th}),
%\end{equation}

\subsubsection{Spike bursts}
\label{spiking}
Slow waves are continually present in GI tissues, but they do not, alone, generate a substantial active response. Mechanically significant contractions, instead, arise due to the occurrence of fast action potentials, termed spikes, that are superimposed onto the slow wave (see Fig.~\ref{generirepr}b) \citep{fung2020functional, kuruppu2021high}. The probability of spike generation is modulated by enteric neural, endocrine, and paracrine inputs that shift the baseline membrane potential of SMC's, thereby facilitating or suppressing local activation. As a result, intestinal wall motility combines a regular background activation wave with episodic, spatially confined over-activation events capable of producing strong localized contractions. These episodes are essential for luminal occlusion, segmental mixing, bolus propulsion, and driving the formation of folds and wrinkles. 

\begin{figure}[h]
\centering\includegraphics[width=\textwidth]{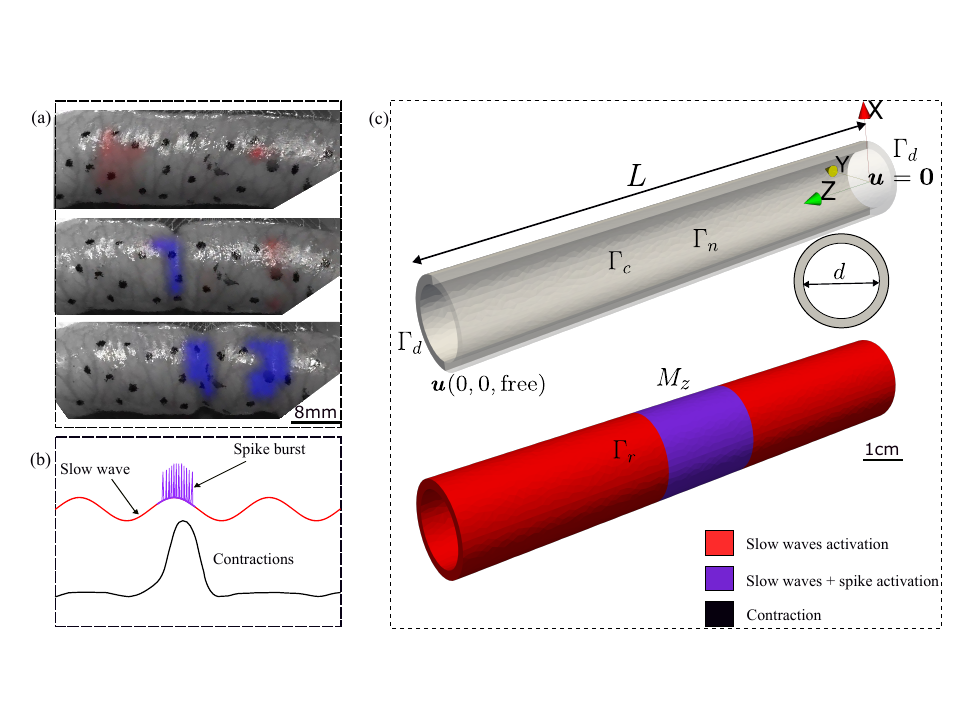}
    \caption{Schematic representation of the problem.
(a) Experimental observation of pig intestine contraction~\citep{kuruppu2021high}: the red color
represents the slow wave, while the blue color represents the spike burst.
(b) Generic representation of the slow wave associated with the spike burst;
red denotes the slow wave, blue denotes the spike burst, and the black curve
shows the resulting contraction wave.
(c) Problem configuration: the geometry is a hollow cylinder with an outer
diameter of 3.0~cm and a wall thickness of 0.3~cm and length 20~cm. One end is fully fixed with $\bu = {\bf 0}$,
while the other end is free to move along the $z$-axis only, i.e. $\bu(0, 0, \rm free)$. The slow wave spatiotemporal activation is present throughout the domain (red), whereas the blue area identifies the spike burst localization. $\Gamma_c$ is the possible contact surface, $\Gamma_r$ accounts for the presence of surrounding organs, $\Gamma_n$ is used to account for the presence of chyme.}
    \label{problem_config}
\end{figure}

Our understanding of such phenomena stems from experimental observations, such as those concerning peristaltic contractions in pig intestines, see Fig.~\ref{problem_config}a \citep{kuruppu2021high, nagahawatte2023high}. %The red signal indicates the propagation of slow waves, while the blue signal represents the spike waves responsible for activating intestinal contraction.
In the present work, we simulate this behavior for the first time. The baseline wave, originating from electrophysiological slow wave propagation, remains unchanged, providing the underlying activation pattern. On top of this structure, we superimpose a localized wave representing the effect of spike-mediated reinforcement and enteric modulation, see Fig.~\ref{problem_config}b. The resulting mathematical relation for the total action potential of the SMC is thus provided by the relation
\begin{equation}
u_s^{\mathrm{tot}}(\bx,t)
=
u_s(\bx,t)
+
u_{s,p}(\bx, t),
\label{slow_spike}
\end{equation}
where $u_s$ denotes the slow wave spatiotemporal dynamics whilst $u_{s,p}$ represents the additional spike-burst activity localized in space, as shown in Fig.~\ref{problem_config}c.

Local activation is subsequently described using two complementary strategies that differ in whether explicit synchronization with the propagating wave is enforced. Specifically, we model spike-bursts by considering an additional component that can either be (i) out of phase with the slow wave, representing the neurogenic case, or (ii) in phase, representing the myogenic case \citep{kuruppu2021high}.

\paragraph{Neurogenic spikes}

The spike space-time model is formulated generally as:
\begin{equation}
    u_{s,p}(\bx,t)= A(t)\,M_z(\bx),
    \label{burst}
\end{equation}
where, $A(t)$ represents the temporal envelope of the activation and $M_z(\bx)$ defines its spatial profile. In this representation, the spike-burst activity is fully decoupled from the slow wave, corresponding to a purely neurogenic spike scenario. 

\paragraph{Myogenic spikes}
Physiologically, spike activity is frequently seen to synchronize with slow waves. We account for this behavior by assuming that successive passages of the slow wave crest can accumulate locally and eventually trigger the occurrence of spike bursts. Accordingly, we introduce an internal state variable $s(\bx,t) \in [0,1]$ such that:
\begin{equation}
u_{s,p}(\bx,t)=A_p\,s(\bx,t),
\label{synchro-spike}
\end{equation}
where $s=s(u_s(\bx,t),M_z(\bx))$ evolves according to the local slow wave intensity. Specifically, it increases when the baseline activation exceeds a given threshold, $\uth$, within the target region $M_z=M_z(\bx)$ and relaxes otherwise, such that
\begin{equation}
\label{eq:sdot}
\frac{d s}{dt}
=
\frac{1-s}{\tau_{\mathrm{on}}}\,T_r
-
\frac{s}{\tau_{\mathrm{off}}}(1-T_r), \quad T_r = H(u_s-\uth)\,M_z,
\end{equation}
where $\tau_{\rm on}$ and $\tau_{\rm off}$ are the activation and deactivation time constants, respectively. As a result, successive passages of the slow wave crest can accumulate and generate a progressively stronger local response prior to relaxation. It should be noted that this approach is also applicable to neurogenic, as well as myogenic spikes (i.e., a transition from myogenic activation, driven by slow wave accumulation, to neurogenic activation associated with persistent spike activity and slow wave desynchronization).\\

%The proposed modeling approach enables the simulation of both independent spike activity (neurogenic spikes) and myogenic spikes that are coupled with the slow wave activity.

The total activation function then combines the baseline slow wave and spike burst effects as:
\begin{equation}
\gamma_{\mathrm{tot}}
=
\min\!\left(\gamma(u_s)+\gamma_p(u_{s,p}),\,\gamma_{\max}\right),
\label{total_act}
\end{equation}
where $\gamma_{\max}$ limits non physiological activation values and $\gamma_{\mathrm{tot}}$ is used inside Eq.~\eqref{eq:2}. Although spike activity can arise either independently or in synchrony with the slow wave, its ability to trigger contraction is known to depend on intracellular calcium dynamics. As such, the activation of $\gamma_p(u_{s,p})$ follows Eq.~\eqref{gammafun} with distinct threshold values $\uth$ thus mimicking baseline activation (see Tab.~\ref{param}).

\subsection{Effective contractilities}
\label{contract_model}
An open question relating to the physiology of intestinal motility concerns the coordination between the longitudinal and circumferential muscle layers, which, together, are responsible for the large deformations observed during peristalsis. Several hypotheses have been proposed in the literature, with some studies suggesting the simultaneous contraction of both muscle layers at similar intensities, whilst others have reported region-dependent or reciprocal contraction patterns, in which the longitudinal layer relaxes during circumferential contraction and vice versa \citep{lammers2002similarities}. Recently, a parametric study investigated the role of circumferential and longitudinal contractilities in the stomach \citep{henke2026electromechanical}. The results revealed that luminal occlusion is primarily driven by high circumferential contractility, whereas longitudinal contractility contributes predominantly to wall thickening. %However, in this study, the contractility parameters were assumed to be constant.

Experimental evidence indicates that slow wave and spike-burst activities generate different levels of contraction, and that contraction intensity depends on both the amplitude and duration of pacing \citep{nagahawatte2022gastric, nagahawatte2023pacing}. In particular, higher stimulation amplitudes and longer stimulation durations have been shown to enhance motor unit recruitment and temporal accumulation, thereby increasing muscle force generation and contraction strength.

Taking these observations into account, namely (i) the dominant role of circumferential contractility and (ii) its time-dependent modulation by spike burst activation, a dedicated model for circumferential contractility is introduced:
\begin{equation}
\alpha_c^{\mathrm{circ}}(\bx,t)
=
\alpha_c(r)
%\left[1+\eta_{ac}\,A(t)\,M_z(\bx)\right].
\left[1+\eta_{c}\,{u_{s,p}(\bx,t)}\right]
\label{circum_act}
\end{equation}
whilst leaving the longitudinal contractility $\alpha_l(r)$ unchanged and $\eta_{c}$ stands for control parameter (see Fig.~\ref{generirepr}).
Accordingly, effective contractility functions are introduced for the longitudinal and circumferential layer,
$G_l = \alpha_l(r)\gamma_{\mathrm{tot}}$, 
$G_c = \alpha_c^{\mathrm{circ}}(\bx,t)\gamma_{\mathrm{tot}}$, respectively, thus leading to the generalized active deformation gradient tensor \eqref{total_act}
\begin{equation} \label{Tola_Fa}
    \mathbf{F}_a 
    = 
    \mathbf{I} - (G_c \bA_c
     + G_l \bA_l)  
    + \gamma_n  \bA_n \,,
\end{equation}  
where $\bA_j=\mathbf{n}_j \otimes \mathbf{n}_j$ stands for the associated structure tensors in the orthonormal direction $j=l,c,n$ (longitudinal, circumferential, and normal) \citep{DJOUMESSI2024116989}. It is worth noticing that this formulation ensures effective contractility functions remain bounded by the physical limit of unity.

%\paragraph{Remark}
%We note that experimental evidence is investigating the possibility of spatial desincronization among circumferential and longitudinal layers, though no clear  \citep{nagahawatte2023evaluation}.
%Available experimental observations suggest that the activation regions are similar for both fiber families, although the magnitude of the generated force may differ \citep{kuruppu2024electromechanical, nagahawatte2023evaluation}.
%Besides, due to different intensity of the force produced by the different layers \citep{henke2026electromechanical}, we focus on the contractility of the circumferential layer, indicating that longitudinal contraction contributes to axial shortening and may increase the overall stiffness of the structure, thereby restricting circumferential motility.

\subsection{Mechanical modeling of the intestinal wall}

In line with typical formulations characterizing the constitutive properties of load-bearing soft tissues, we model the intestinal wall as an anisotropic hyperelastic quasi-incompressible material. The tissue is reinforced by four families of fibers with specific preferential directions (see Fig.~\ref{generirfiber}). Accordingly, the total strain energy density $\Psi$ is additively decomposed into isotropic, $\Psi^{\rm iso}$, anisotropic, $\Psi^{\rm aniso}$, and volumetric,  $\Psi^{\rm vol}$, parts. The isotropic matrix material is modeled as a Neo-Hookean material, whilst the anisotropic contribution is modeled as an Ogden-Holzapfel structure-based energy density, with both passive and active terms~\citep {sokolis2021variation,holzapfel2002nonlinear, sokolis2013microstructure}. The passive terms are associated with the mechanical response of directional collagen fibers $(d_1, d_2)$, whilst the active terms are associated with the longitudinal $(l)$ and circumferential $(c)$ muscle layers, such that:
\begin{linenomath}
\begin{equation} 
\label{eq:consti}
  \Psi =
  %\Psi^{\rm iso} +
  %\Psi^{\rm aniso}+
  %\Psi^{\rm vol}
  %= 
  \mu(I_1 -3)+
  \sum_{i \in \{ l, c, d_1, d_2 \}} \frac{\eta_i k_1^i}{4k_2^i}[e^
  {k_2^i(I_4^i -1)_+^2}-1]
  -p \, {\rm log}J + \frac{1}{2\kappa} p^2 
  .
\end{equation}
\end{linenomath}
Here, the notation $(y)_+:= y$ if $y\geq 0$ reproduces the tension-compression switch approximation and the anisotropic fourth invariant $I_4^i = \bC:\bA_i$ is distinguished for each fiber family $i \in \{l,c,d_1,d_2 \}$. The material parameters  $k_1^i$ (stiffness-like) and $k_2^i$ (nondimensional) are associated with the directional behavior of the material, $\eta_i$ modulates fiber stiffness (used for controling the numerical stability), $\mu$ is the passive isotropic stiffness, and $\kappa$ is a stiffness parameter akin to the bulk modulus (see Table~\ref{table:mechanical} for material parameters).

\paragraph{Remark}
The constitutive model excludes an explicit radial contribution, preventing direct control of radial deformation. Enforcing strict incompressibility ($J=1$) would thus artificially constrain thickness changes. A quasi-incompressible formulation is therefore adopted, allowing for physiologically realistic radial deformation.\\

Accordingly, the first Piola-Kirchhoff stress tensor reads:
\begin{equation} \label{Piola}
    \bP 
    = 
    \frac{\partial \Psi}{\partial \bF} 
    {-pJ\bF^{-T} }
    = 
    {\rm det} (\bF_a) \left(\frac{\partial \Psi^{ \rm iso}}{\partial \bar{\bF_e}}+ \frac{\partial \Psi^{ \rm aniso}}{\partial \bar{\bF_e}}\right) \bF_a^{-T}
    -
    pJ\bF^{-T} \,.
\end{equation}

\paragraph{Heterogeneous fiber reinforcement}
An accurate representation of the tissue's fibrous structure is critical to its proper constitutive characterization. Due to the lamellar organization of the GI wall and the strong adhesion between its layers,
%, the thickness of the geometry was discretized with two finite elements. Accordingly, 
muscle fibers are selectively activated within their respective layers through the introduction of energy-based activation functions, $f_c(r)$ and $f_l(r)$, that switch between different fiber families across the wall thickness:
\begin{linenomath}
\begin{subequations}
\begin{align}
    f_c (r) &=  c_{\text{base}} + c_{\text{cir}} \left(1 - H(r - r_c)\right) \,, 
    \label{fun_c}
    \\
    f_l (r) &= l_{\text{base}} + l_{\text{long}} \left(1 - H(r_c - r)\right) \,.
    \label{func_c}
\end{align}
\label{function}
\end{subequations}
\end{linenomath}
Here, $c_{\text{base}}$ and $l_{\text{base}}$ denote baseline values to avoid null material parameters; $H(\cdot)$ is a Heaviside step function with parameter $r_c$ controlling the relative thickness of the two layers (see Fig.~\ref{generirepr}a). Accordingly, mechanical energy parameters are defined 
$k_i^c(r) = f_c(r)\,k_i^c$,
$k_i^l(r) = f_l(r)\,k_i^l$ for the circumferential and longitudinal layer, respectively, with 
$i \in \{1,2\}$. Distribution functions are provided in Tab.~\ref{newMechParam}.
%The distribution functions are specified with $c_{\mathrm{cir}} = l_{\mathrm{long}} = 1$, and $c_{\mathrm{base}} = 10^{-3}/k_i^c$, $l_{\mathrm{base}} = 10^{-3}/k_i^l$. The same approach is used to computed $\alpha_c(r)=f_c(r)\alpha_c$ and $\alpha_l(r)=\alpha_l(r)f_l(r)$. 

\begin{table}[h!]
    \centering
    \caption{Mechanical parameter based on layer activation functions \citep{brandstaeter2018computational}.}
    \begin{tabular}{c c c c c c}
    \hline
     $c_{\mathrm{cir}}$  & $l_{\mathrm{long}}$ & $c_{\mathrm{base}}$ & $l_{\mathrm{base}}$ & $\alpha_c(r)$ & $\alpha_l(r)$\\
     \hline
        $1.0$ & $1.0$ & $10^{-3}/k_i^c$ & $10^{-3}/k_i^l$ & $\alpha_c^{\mathrm{circ}}f_c(r)$ & $\alpha_l(r)f_l(r)$\\
    \hline
    \end{tabular}
    \label{newMechParam}
\end{table}

\ref{param} provides the full list of material parameters and a representative distribution of through-the-thickness mechanical parameters and contractility functions.

\subsection{Contact Model}
\label{sec:contactMod}
Our frictionless contact formulation relies on a variational augmented Lagrangian approach, allowing the impenetrability condition to be weakly enforced without explicitly introducing inequality constraints \citep{renard2013generalized, poulios2015unconstrained}. The formulation is particularly well suited for large deformations, large sliding, and self-contact problems and relies on a ray-tracing strategy to map contact points between deformable surfaces.

Let $\bx = \phi(\bX)$ denote a spatial point on the  intestine luminal surface $\Gamma_c$ (see Fig. \ref{problem_config}) and let $\by = \phi(\bY)$ be its associated point on the same intestine surface $\Gamma_c$, determined via ray-tracing along intestine surface normal vector $\bn_\bx$. The gap function, measuring interpenetration is defined as:
\begin{equation}
   g(\bX) = \bn_\bx \cdot (\by - \bx)
   \label{gap_ray}
\end{equation}
where, $\bn_\bx$ is the outward unit normal to the deformed intestine luminal surface at point $\bx$ (see \citep{poulios2015unconstrained} for the definition). By considering only one contact surface, the non-penetration condition $g(\bX) \geq 0$ is imposed through a Lagrange multiplier $\lambda_N(\bX) \leq 0$ where $\lambda_N : \Gamma_c\rightarrow \mathbb{R}$ represents the material normal contact traction.

To avoid handling inequality constraints explicitly, the problem is reformulated using an augmented Lagrangian functional:
\begin{equation}
    \mathcal{L}_r(\bu, \lambda_\bN) = \mathbb{P}(\bu, p) + \frac{1}{2r_p} \int_{\Gamma_c^S} [\lambda_N + r_p g]_{-}^2 - (\lambda_N)^2 \, d\Gamma,
    \label{lag_f}
\end{equation}
where, $\mathbb{P}(\bu, p)$ is the total potential energy (including internal elastic energy and external loading), $r_p > 0$ is the  augmentation parameter (stiffness-like), and $[a]_{-} = \min(x, 0)$ denotes the negative part of $a$. The set $\Gamma_c^S \subset \Gamma_c$ identifies the surface region where contact is detected. The augmented Lagrangian parameter is particularly challenging to tune in the context of large deformations involving fiber-reinforced hyperelastic tissues. To estimate an appropriate value, we assumed that \( r_p \) scales with the average of \( \mu \) and \( k_1^c \), such that
\begin{equation}
r_p = \varsigma \, \frac{\mu + k_1^c}{2},
\end{equation}
where $\varsigma$ is a dimensionless proportionality coefficient.

\section{Model exploitation}
\label{application}

The numerical methodology, including the derivation of the weak formulation, the contact algorithm, and the details of the numerical implementation, is presented in Appendix~\ref{numerical}. The following section therefore focuses on the exploitation of the proposed model and the analysis of the resulting numerical simulations

\subsection{Pre-stress configuration}
Before starting any simulations, it is important to note that, the computational geometry was constructed based on the geometrical dimensions measured from the experimental intestinal specimen \citep{kuruppu2021high}, which was subjected to mild luminal air inflation during acquisition. This implies that the reference geometry already reflects a physiologically prestressed state, rather than a truly stress-free configuration.
To explicitly incorporate this prestressed state into the computational framework, a uniform luminal pressure was applied to the reference cylindrical mesh (see Fig.~\ref{problem_config}c) prior to any electromechanical activation. This inflation step serves to mechanically consistent the reference configuration with the in vivo conditions of the tissue. The resulting residual von Mises stress field, on the order of $\sim 0.5$ kPa, confirms the existence of a non-trivial internal stress state embedded in the reference configuration. Furthermore, the negligible geometric change observed during this inflation step validates that the constructed mesh already closely approximates the prestressed in vivo configuration, ensuring full consistency between the computational model and the experimental measurements used for validation.

\subsection{Spatial activation profiles}
Before presenting the numerical simulations, it is important to justify the choice of the spatial activation profiles adopted to represent spike-burst activity. Unlike slow waves, which are primarily generated by interstitial cells of Cajal (ICC) and can be mapped with high spatial resolution, spike-bursts reflect the electrical excitability of smooth muscle cells (SMCs) and are associated with elevations in intracellular calcium that induce contraction. Such activity can be facilitated or inhibited by neurotransmitters released by the enteric nervous system (ENS). However, the mechanisms underlying spike-bursts remain particularly difficult to characterize directly \emph{in vivo}. This limitation arises from the complex interplay among ICC, SMCs, and enteric neurons within the intestinal wall, as well as the continuous motion of the intestine and the low amplitude of the recorded signals, all of which, despite recent advances in high-resolution mapping techniques, complicate the measurement of intestinal electrophysiology \citep{boys2025implantable, cheng2013mapping, lammers1993multielectrode}. Consequently, the available experimental studies provide only indirect information on spike-burst activity through the resulting intestinal contractions. Careful examination of experimental data \citep{kuruppu2024electromechanical} reveals that the spatial distribution of intestinal contractions varies significantly from one spike-burst event to another. While some contractions exhibit a localized and nearly symmetric pattern, others display markedly asymmetric distributions or abrupt spatial transitions. 

In the present modeling framework, we do not attempt to reproduce each experimental event individually but select three spatial activation profiles, namely Gaussian, Heaviside, and hybrid Gaussian--Heaviside (see \ref{sec:B}) to reproduce the different motility patterns as discussed in Sec.~\ref{sec:optimal_control}.
In this section, we present an idealized case study discussing the effect of spike burst and their superposition with slow waves excitation assuming a Gaussian spatial activation profile symmetrically identified along the computational domain as shown Fig.~\ref{problem_config}.

\subsection{Neurogenic spike burst}
We start considering the localized spike activation as an externally prescribed stimulation independent of the baseline slow waves activation.
Eq.~\eqref{burst} characterizes as Gaussian profile
with a sinusoidal time modulation
$A(t) = A_m \sin\left(\pi {(t - t_{\mathrm{on}})}/{(t_{\mathrm{off}} - t_{\mathrm{on}})}\right)$.
Spike activation times are defined
$
t_{\rm on} = t + t_{\rm delay}, t_{\rm off} = t_{\rm on} + t_{\rm spi\_dur},
$
where $t_{\rm delay}$ is the spike triggering delay and $t_{\rm spi\_dur}$ the spike activation time duration. 
%This formulation allows precise control of when the spike-induced contraction starts and ends at a given location, thereby influencing the effective recruitment motor unit and the resulting deformation patterns in the intestinal wall. 
%Modulating $t_{\rm delay}$ and $t_{\rm spi\_dur}$, different physiological scenarios can be simulated.

\begin{figure}[h]
    \centering

    \begin{subfigure}{0.46\textwidth}
        \includegraphics[width=\linewidth]{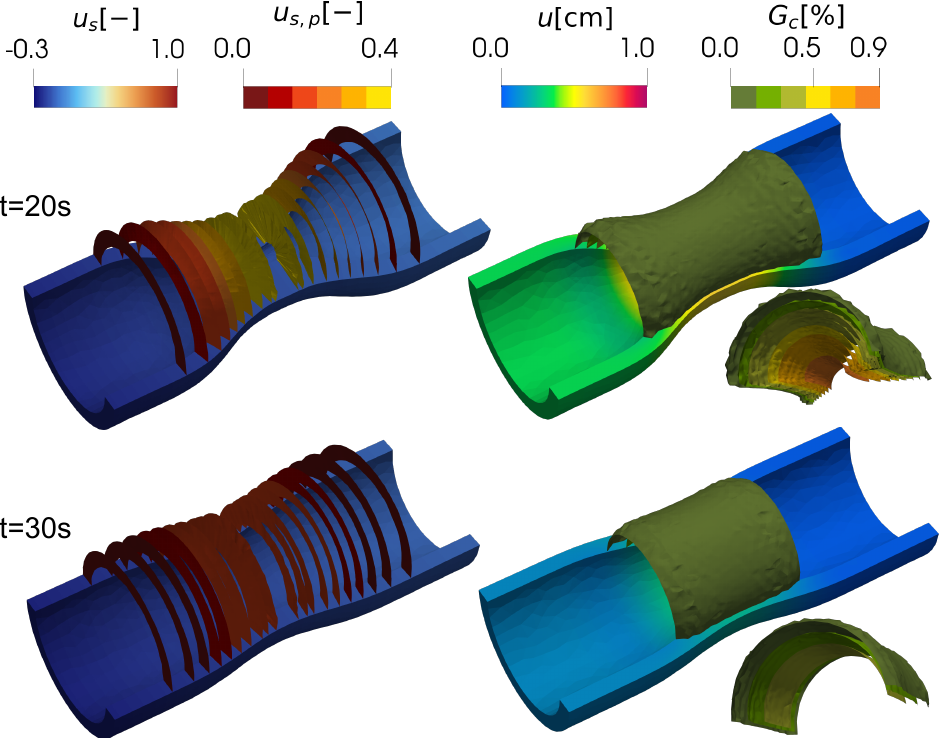}
        \caption{}
        \label{fig:non_syncro_a}
    \end{subfigure}
    \hfill
    \begin{subfigure}{0.46\textwidth}
        \includegraphics[width=\linewidth]{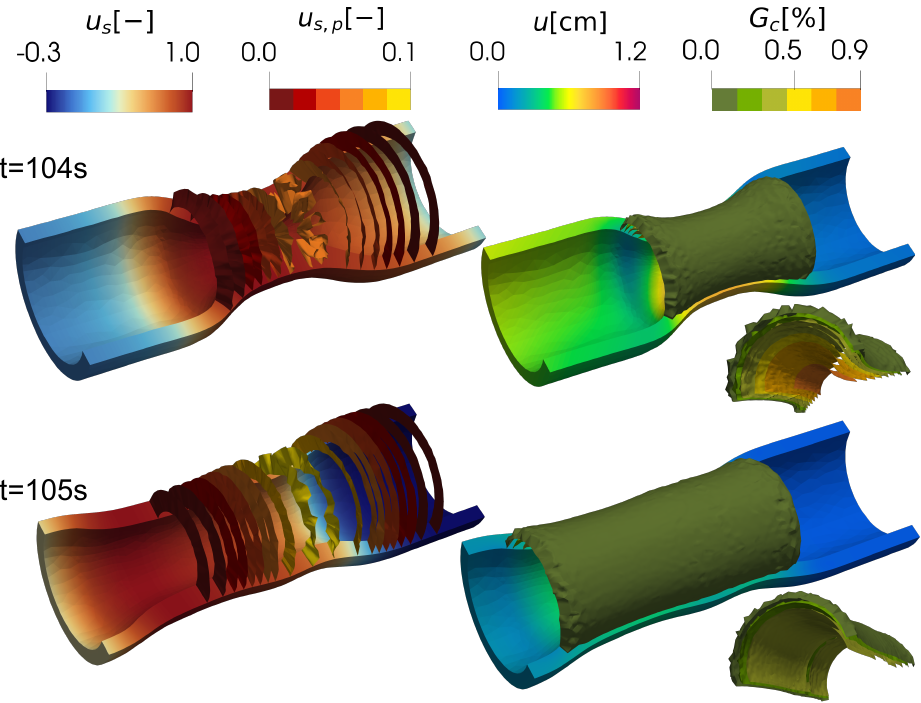}
        \caption{}
        \label{fig:non_syncro_b}
    \end{subfigure}
    \caption{Temporal and spatial evolution at two representative time frames of (a) neurogenic and (b) myogenic spike burst. Each panel provides (left) slow wave membrane voltage $u_s$ superimposed with the iso-contour of the spike burst voltage $u_{s,p}$, and (right) the displacement magnitude $u$ superimposed with the iso-contour of the effective contractility $G_c$. Inset is provided as isometric iso-contour view of $G_c$. Model parameters $z_0 = 10\,\mathrm{cm}$, $\sigma_z = 1.2$, $\tau_{\mathrm{on}} = 8$, $A_p = 0.45$, $u_{\mathrm{tr}} = 0.2$, and $\eta_{c}=1.5$.
    }
    \label{non_syncro}
\end{figure}

Figure~\ref{non_syncro}a shows numerical results superimposing SMC slow wave membrane potential, $u_s$, with spike burst contours, $u_{s,p}$, and mechanical displacement magnitude, $u$, with  effective contractility contours, $G_c$, for two time frames. As expected, the additional localized spike burst field activity remains desynchronized from the slow waves spatio-temporal propagation, thus contributing to mechanical activation only when overlapping to the spike burst region. Beyond the ability to reproduce peristaltic (PS) deformation, the strong localized contraction leads to lumen occlusion reaching a physiological reduction of $80\%$ when the effective contractility $G_c$ reaches its maximum of $90\%$ and the slow wave passes on the same region. 

%This type of motility has been experimentally identified as segmental or spontaneous peristaltic contraction. Here we quantify for the first time this phenomenon comparing the characteristic length of the contractile zone with available experimental data \citep{kuruppu2024electromechanical}. Figure~\ref{wrin}a confirms a very good estimate of $\sim1.6$~cm. 

\subsection{Myogenic spike burst}
\label{myogenic spike burst}
In this case, the activation is modeled such that spike burst activity is temporally synchronized with the slow wave (and still spatially co-localized in the spike burst region), as described by Eqs.~\eqref{synchro-spike}--\eqref{eq:sdot}. 
%The spatial distribution of the activation follows the same Gaussian profile as that used in the neurogenic case.
%
Figure~\ref{non_syncro}b shows the superimposed fields comparison highlighting that the model predicts a propagating peristaltic contraction (at $t=20\,s$) but also, for the first time in an active electromechanics of gastrointestinal motility, the emergence of wrinkling (WK) patterns (at $t=105\,s$).
The mechanistic origin of this behavior can be attributed to the close spatial proximity (approximately \(2\,\mathrm{cm}\)) between the slow wave activation front and the spike burst triggering region, combined with the imposed layered heterogeneity on the intestinal wall thickness. Under these conditions, neighboring regions experience different activation states and mechanical loads, leading to localized mechanical instabilities that manifest as wrinkling patterns.

\subsection{Characteristic lengths}
\begin{figure}[h]
    \centering

    \begin{subfigure}[t]{0.48\textwidth}
        \centering
        \includegraphics[width=\linewidth]{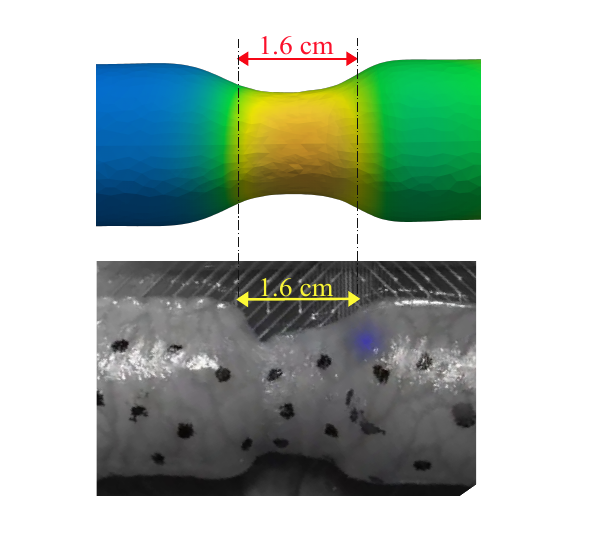}
        \caption{}
        \label{fig:nw1}
    \end{subfigure}%\hspace{0.01\textwidth}
    \begin{subfigure}[t]{0.48\textwidth}
        \centering
        \includegraphics[width=\linewidth]{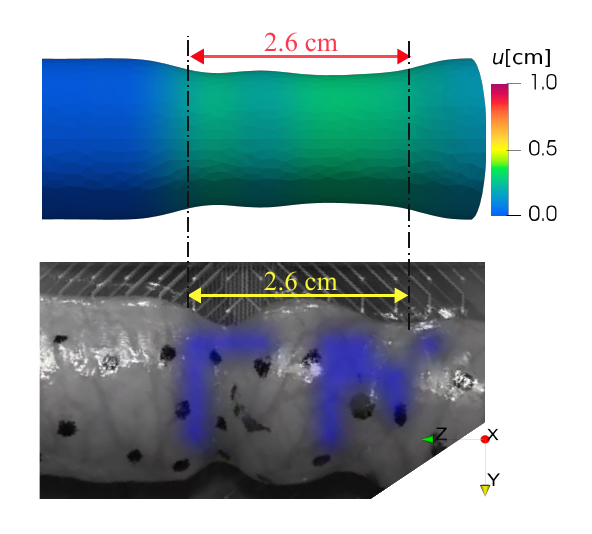}
        \captionsetup{margin={-1.1cm,0cm}}
        \caption{}
        \label{fig:nw2}
    \end{subfigure}

    \caption{Quantitative and qualitative validation of the apparent contraction lengths. Comparison between simulated and experimental data \citep{kuruppu2024electromechanical} in the case of peristalsis (a) and wrinkling (b).}
    \label{wrin}
\end{figure}

The previous analyses showed that the neurogenic case is characterized exclusively by peristaltic contractions, whereas the myogenic case generates both peristalsis and wrinkling patterns. This observation raises an important question regarding the underlying mechanisms that differentiate the PS and WK responses. As preliminary validation, we quantify the predicted characteristic lengths for the two regimes comparing and contrasting simulation patterns with experimental data \citep{kuruppu2024electromechanical} in Fig.~\ref{wrin}. 

%To the best of our knowledge, this is the first study to quantitatively compare the characteristic lengths of both peristaltic and wrinkling regions obtained experimentally with those predicted by an electromechanical model.
%
For the peristaltic regime, both the neurogenic and myogenic simulations yield a characteristic length of $\sim1.6$~cm, which is in excellent agreement with experimental observations (Fig.~\ref{wrin}a).
In the case of wrinkling (simulated for myogenic excitation only), a characteristic length of approximately $\sim2.6$~cm is obtained resulting in good agreement with available experimental measurements (Fig.~\ref{wrin}b).

\begin{comment}
    \begin{figure}[h]
    \centering
    \includegraphics[width=0.8\linewidth]{new_wrikling4.pdf}
    \caption{Quantitative and qualitative validation of the apparent contraction lengths. Comparison between simulated and experimental data \citep{kuruppu2024electromechanical} in the case of peristalsis (a) and wrinkling (b).}
    \label{wrin}
\end{figure}
\end{comment}

\subsection{Impact of fiber segregation on lumen occlusion}
\begin{figure}[h]
    \centering

    \begin{subfigure}{0.35\textwidth}
        \includegraphics[width=\linewidth]{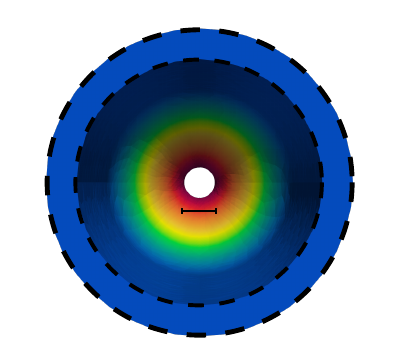}
        \caption{Layered}
        \label{comparison_a}
    \end{subfigure}
    %\hfill
    \begin{subfigure}{0.35\textwidth}
        \includegraphics[width=\linewidth]{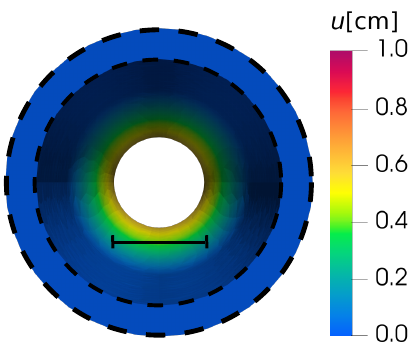}
        \captionsetup{margin={-1.1cm,0cm}}
        \caption{Homogenized}
        \label{comparison_b}
    \end{subfigure}
    \caption{Comparison of intestinal lumen displacement magnitude, $u$, highlighting lumen occlusion obtained with the layered (a) and homogenized \citep{DJOUMESSI2024116989} (b) muscle fiber architectures within intestinal wall. The layer-specific formulation produces lumen occlusion approaching $80\%$ (diameter $0.47 \rm cm$), compared with the homogenized model that reaches $40\%$ occlusion (diameter $1.3 \rm cm$). Black dashed lines identify the reference configuration.}
    \label{comparison}
\end{figure}

In the previous study, we observed contraction levels that were sufficiently large to nearly achieve lumen occlusion (we recall the contact model in Sec.~\ref{sec:contactMod} is solved during electromechanical simulations). In an idealized geometry, this behavior totally depends on the layer-specific representation adopted, i.e., circumferential and longitudinal muscle fibers assigned to their respective anatomical layers. Such a strategy avoids the artificial mechanical competition that arises when both fiber families are constitutively homogenized throughout the intestine wall~\citep{DJOUMESSI2024116989}. Figure~\ref{comparison} illustrates, for a given electrical stimulation, the lumen occlusion predicted by the present model (a) and the corresponding response obtained with the homogenized formulation (b) at the time of spike burst activation. A substantial difference can be observed: the predicted lumen occlusion reaches $\sim30\%$ in the homogenized case, which increases by $\sim50\%$ in a realistic layered model, leading to an overall closure of $\sim80\%$.

\paragraph{Remark} A layer-specific constitutive modeling approach allows the circumferential layer to contract more efficiently, resulting in substantially enhanced lumen occlusion in agreement with experimental and clinical evidence.

\newpage
%\clearpage

\section{Critical condition for peristalsis (PS) and wrinkling (WK)}
\label{sec:optimal_control}
This section investigates in detail the mechanisms responsible for the onset of peristaltic and wrinkling regimes according to the selected spatial activation profile. 
%To this end, we first analyze the myogenic case and then introduce an alternative spatial distribution of the spike-burst activity to assess its influence on the resulting mechanical response.

\begin{figure}[h]
    \centering
    \includegraphics[width=0.4\linewidth]{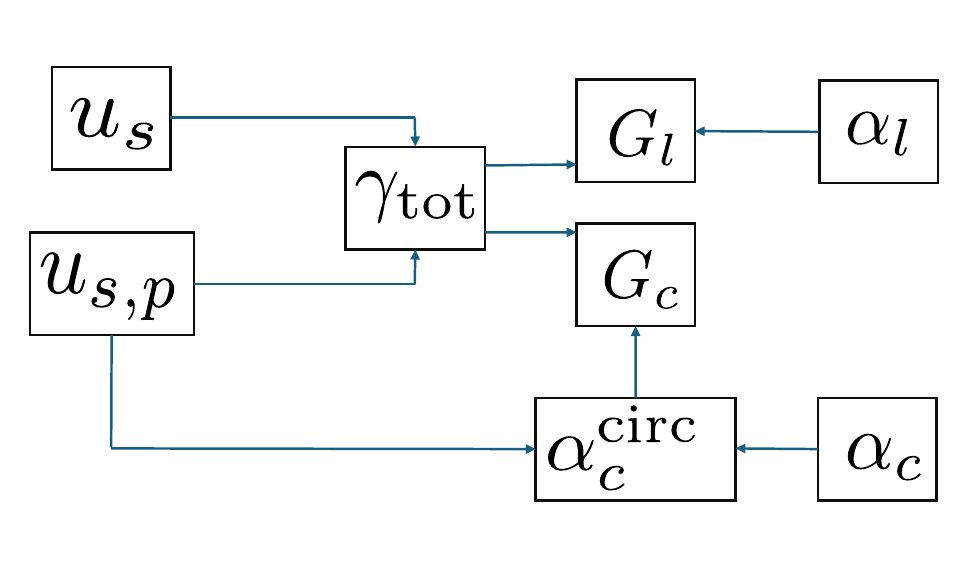}
    \caption{Schematic representation of the interactions between the electrophysiological fields, activation variables, and effective contractions. The flowchart highlights how the slow wave field $u_s$, spike-burst field $u_{s,p}$, and activation variables $\alpha_c$ and $\alpha_l$ contribute to the effective circumferential and longitudinal contractilities ($G_c$ and $G_l$), ultimately governing the emergence of peristaltic and wrinkling patterns.}
    \label{flowchart}
\end{figure}

To facilitate the interpretation of the results, Fig.~\ref{flowchart} presents a flowchart summarizing the coupling between the different model quantities. The transmembrane voltage state variables $u_s $ and $u_{s,p}$, together with the layer dependent activation variables $\alpha_c$ and $\alpha_l$, constitute the main control variables of the framework, while the effective circumferential and longitudinal contractilities, $G_c$ and $G_l$, respectively, represent the resulting outputs governing tissue contraction.

%\clearpage
%\newpage
\subsection{Gaussian profile}

\begin{figure}[!h]
\centering\includegraphics[width=\textwidth]{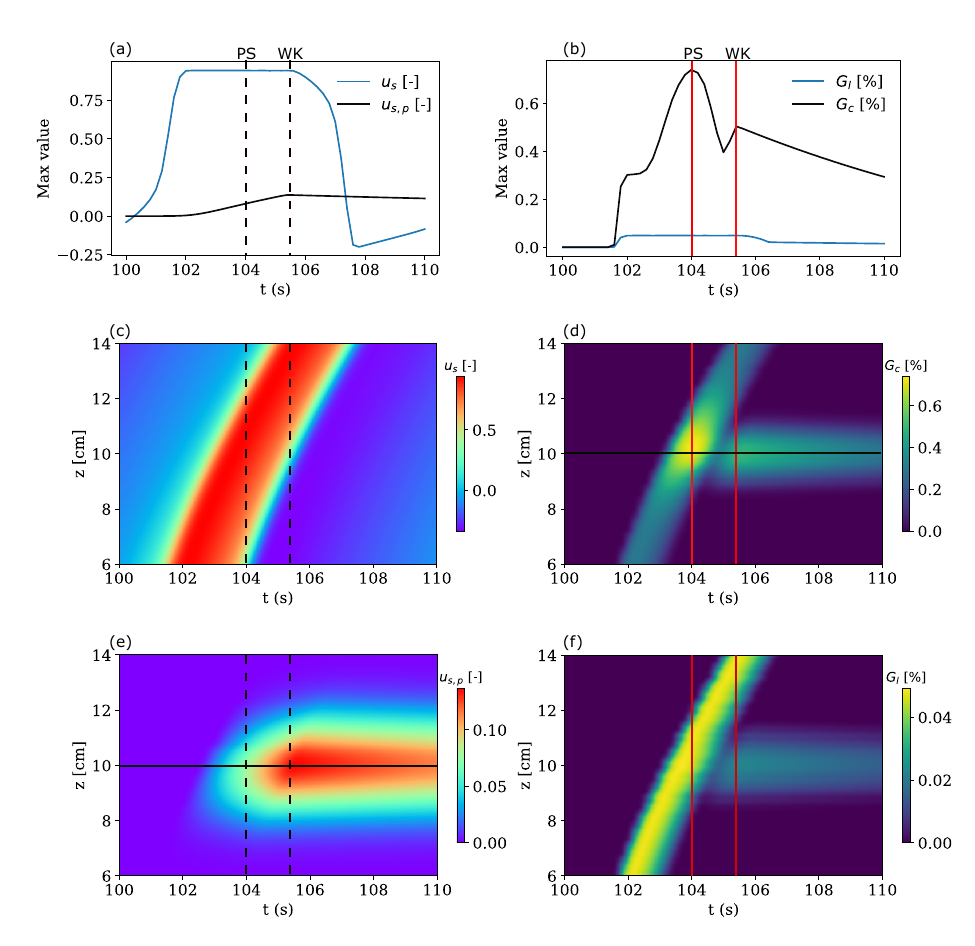}
    \caption{Gaussian profile of spike burst activation function. Space-time analysis of the interaction between slow waves, spike-burst activity, and muscle contraction. (a) Temporal evolution of the maximum amplitudes of the slow wave $u_s$ and spike-burst $u_{s,p}$ activity. (b) Maximum effective contractility of the longitudinal $G_l$ and circumferential $G_c$ layer.  (c-f) Spatio-temporal distributions of $u_s$, $G_c$, $u_{s,p}$, and $G_l$ extracted along the selected probes (see \ref{sec:B}). Vertical lines identify the PS and WK regime. Horizontal lines identify the symmetry of $u_{s,p}$ and $G_c$ in space.} 
     \label{comparative_gauss}
\end{figure}

The Gaussian profile represents the myogenic case discussed previously. Here, we perform an extended quantitative analysis to identify the conditions associated with each deformation regime considering several quantities extracted along a circumferential and longitudinal probe line (see \ref{sec:B}):
the maximum values of the slow wave field $u_s$ and the spike-burst field $u_{s,p}$ (Fig.~\ref{comparative_gauss}a);
the maximum effective circumferential and longitudinal contractilities, $G_c$ and $G_l$, respectively (Fig.~\ref{comparative_gauss}b);
the spatio-temporal evolution of $u_s$, $G_c$, $u_{s,p}$, $G_l$ (Figures~\ref{comparative_gauss}c--f).   

We highlight PS and WK regimes by vertical lines that indicate their onset time.

\subsubsection{PS regime}
Within the PS regime, a strong localized contraction develops. This contraction occurs when the slow wave activity $u_s$ reaches its maximum while the spike-burst activity $u_{s,p}$ increases. Under these conditions, the circumferential layer generates a strong contraction, whereas the longitudinal one retracts sufficiently to accommodate the deformation (in our simulation it happens around $t=104$~s). This mechanism is reflected by the simultaneous maxima of the effective contractilities $G_c$ and $G_l$. The substantial reduction of the effective longitudinal stiffness locally facilitates the circumferential contraction and promotes the development of a localized strong peristalsis.

A distinctive feature of the PS regime can be identified in the space--time diagrams. The vertical line associated with the PS state intersects the pattern at a single spatial location, indicating that the contraction remains spatially localized.

%see \ref{gauss} for a 3D representation

\paragraph{Remark}
We emphasize that an efficient peristaltic response relies on the coordinated action of the circumferential and longitudinal muscle layers. Specifically, longitudinal shortening alleviates the mechanical constraint acting on the circumferential layer, thereby reducing the mechanical competition between the two muscle layers. Consequently, the circumferential contraction can develop more efficiently, leading to a more pronounced lumen occlusion.

%
%

%However, unlike the \textit{PsW} regime shown in Fig.~\ref{comparative_brute}, no coexistence between strong contraction and slight wrinkling is observed here. This observation reinforces the hypothesis that the simultaneous appearance of wrinkling strongly depends on the local spatial asymmetry of the spike-burst wave. Indeed, Figs.~\ref{comparative_gauss}c and f show a pronounced spatial symmetry of both the spike-burst activity $u_{s,p}$ and the effective contractility $G_c$, whose axis of symmetry is located around $z=10$ cm. This spatial symmetry appears to limit the emergence of wrinkling instabilities during strong contraction phases.  

\subsubsection{WK regime}
The WK regime exhibits a markedly different behavior. The emergence of wrinkling appears to be governed by two main mechanisms.
First, the circumferential contractility progressively loses its dominance as it relaxes, while the contribution of the longitudinal layer becomes increasingly significant (in our simulation at $t=105$~s). As a result, the balance between the two muscle layers changes, reducing the ability of the circumferential layer to generate a localized peristaltic contraction and promoting a more complex mechanical response.
Second, the slow wave and spike-burst activities become locally desynchronized in space. As observed in Figs.~\ref{comparative_gauss}d and f, a vertical line representing a given instant in time intersects the activation patterns at two distinct spatial locations, approximately $z=10$~cm and $z=12$~cm. Consequently, two neighboring regions of the intestine experience different activation states simultaneously.

This coexistence of spatially close activation zones induces simultaneous loading of the circumferential and longitudinal fibers in neighboring regions. Consequently, the tissue is subjected to non-uniform mechanical loading. Coupled with the intrinsic heterogeneity of the material properties, such conditions favor local deformation incompatibilities and trigger mechanical instabilities manifested as wrinkling patterns. This mechanism is further reflected in Fig.~\ref{comparative_gauss}b, where a bifurcation-like feature appears.

%\clearpage
%\newpage

\subsection{Hybrid Gaussian--Heaviside profile}
\begin{figure}[ht]
\centering\includegraphics[width=\textwidth]{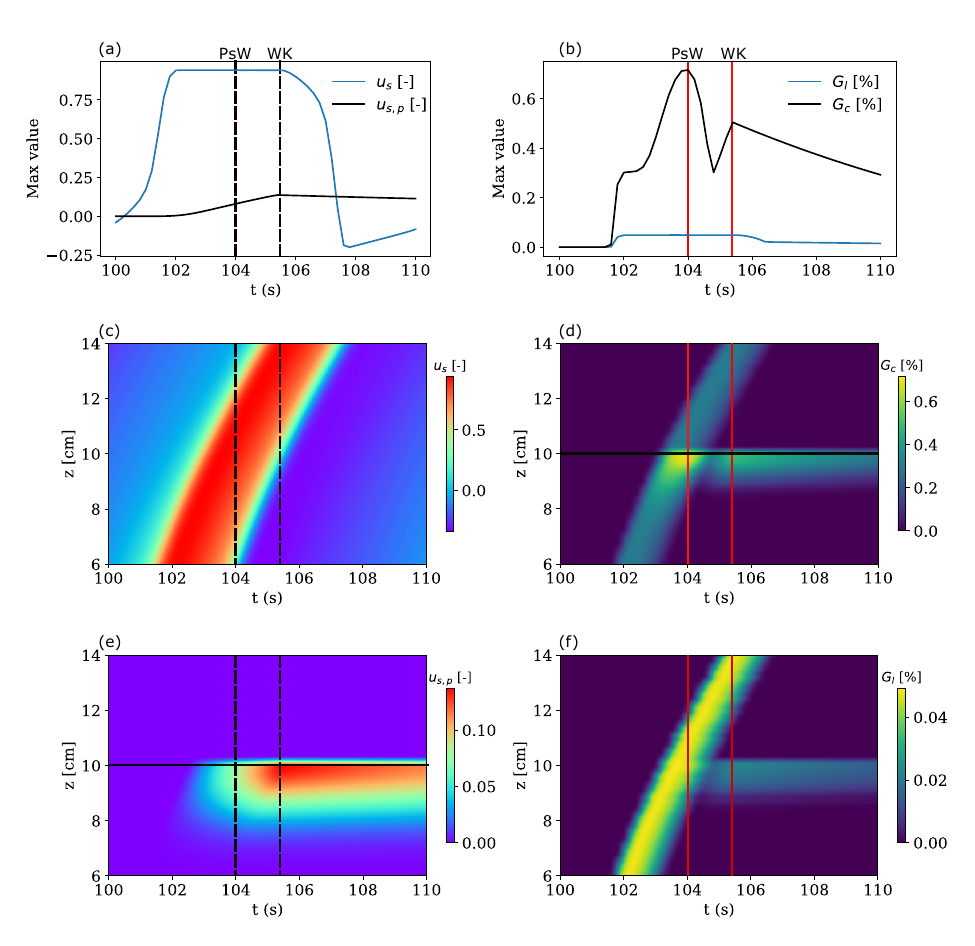}
   \caption{Hybrid Gaussian--Heaviside spike burst activation profile. Space-time analysis of the interaction between slow waves, spike-burst activity, and intestinal muscle contraction. (a) Temporal evolution of the maximum amplitudes of the slow wave $u_s$ and spike-burst $u_{s,p}$ activity. (b) Maximum effective contractility of the longitudinal $G_l$ and circumferential $G_c$ layer.  (c-f) Spatio-temporal distributions of $u_s$, $G_c$, $u_{s,p}$, and $G_l$ extracted along the selected probes. Vertical lines identify PsW and WK regimes. Horizontal lines identify the non-symmetric $u_{s,p}$ and $G_c$ spatial distribution.} 
     \label{comparative_brute}
\end{figure}

Having identified the mechanisms governing the PS and WK regimes, we next analyze the sensitivity of these conditions considering the hybrid Gaussian--Heaviside spatial activation profile. 
%For this purpose, an additional activation profile is introduced, defined as a combination of a half-Gaussian function and a Heaviside step function, with the parameter $z_c$ controlling the transition between the two regions (see \ref{sec:B}):
%\begin{equation}
%M_z(\bx)=\exp\!\left(-\frac{(z-z_0)^2}{2\sigma_z^2}\right)\,H(z_c - z)
%\label{Gaussian_heavi_mask}
%\end{equation}
In this case, a slight wall deformation is observed during the peristaltic contraction followed by a more pronounced wrinkling. Figure~\ref{comparative_brute} highlights these two regimes, identified by vertical time-lines:  
(i) a new PsW regime, corresponding to the coexistence of peristalsis (strong contraction) and slight wrinkling;  
(ii) a WK regime, characterized by wrinkling alone. 
%(see Fig.~\ref{result_br} for a graphical representation).  

The conditions associated with each regime are shown in Figs.~\ref{comparative_brute}(a-f), confirming that the development of a strong peristaltic contraction requires the circumferential contractility to remain dominant. Nevertheless, a slight wrinkling pattern observed within during PsW indicates that, this wrinkling, in general, does not originate from a spatial desynchronization of the activations but it appears to be induced by the spatial asymmetry of the spike-burst activation profile, which directly affects the distribution of the effective circumferential contractility. Such a nonlinear phenomena can be identified in Figs.~\ref{comparative_brute}(d) and (e), where the horizontal line highlights the loss of symmetry in the contractility distribution.

As a consequence, the muscle fibers are simultaneously subjected to two distinct contraction modes, one associated with the Gaussian component and the other with the Heaviside component of the spike-burst profile. The resulting non-uniform loading generates a local mechanical imbalance that manifests as a geometric instability in the form of wrinkling.

\paragraph{Remark}
The introduction of an additional spike-burst activation profile constructed from a Heaviside step function (see ~\ref{sec:B} and ~\ref{heavisidefunction}), despite its markedly different spatial distribution, does not produce any localized wrinkling. This observation further supports the hypothesis that the wrinkling observed in the PsW regime originates primarily from the spatial asymmetry of the spike-burst activation, rather than from its specific functional form.

%\clearpage
%\newpage
\section{Assessment of lumen occlusion}
\label{quantitative_r}

Previous analyses revealed that spatial asymmetries in spike burst activation can induce peristalsis and wrinkling-like patterns. Nevertheless, the ultimate physiological quantity of interest, namely the degree of intestinal lumen occlusion, needs careful consideration. 

\subsection{Role of spike burst spatial distribution}
We start by quantifying the effect of spatial activation profile on lumen occlusion. As quantity of interest, we extract the radial strain 
$E_{rr} = \be_r^T\bE \be_r$, with
$\be_r=\left(\nicefrac{x}{r}, \nicefrac{y}{r}\right)$,
$\bE=\nicefrac{1}{2}(\bC-\bI)$,
thus computing the corresponding lumen radius reduction percentage 
$LR=(1-\sqrt{2\bE_{rr}+1})$ and
comparing this value with experimental data \citep{kuruppu2024electromechanical}. In particular, optical transverse maps obtained on pig small intestine are used to assess the deformation measured at the surface in a direction perpendicular to the longitudinal axis thus reflecting the apparent reduction in the external width of the intestinal segment during excitation. %Although obtained from surface-based video analysis, this unique measurement serves as key indicator of circumferential motility, while internal circular contractions manifest as a transverse surface narrowing.

%transverse maps computed for the cases of peristaltic contraction and wrinkling. Figure.~\ref{quantitative}a corresponds to the case of a Gaussian spike burst, Fig.~\ref{quantitative}b to the Gaussian--step activation, and Fig.~\ref{quantitative}c to the smooth step activation. Each panel presents both the transverse strain map and the corresponding lumen.

\begin{figure}[h]
\centering\includegraphics[width=\textwidth]{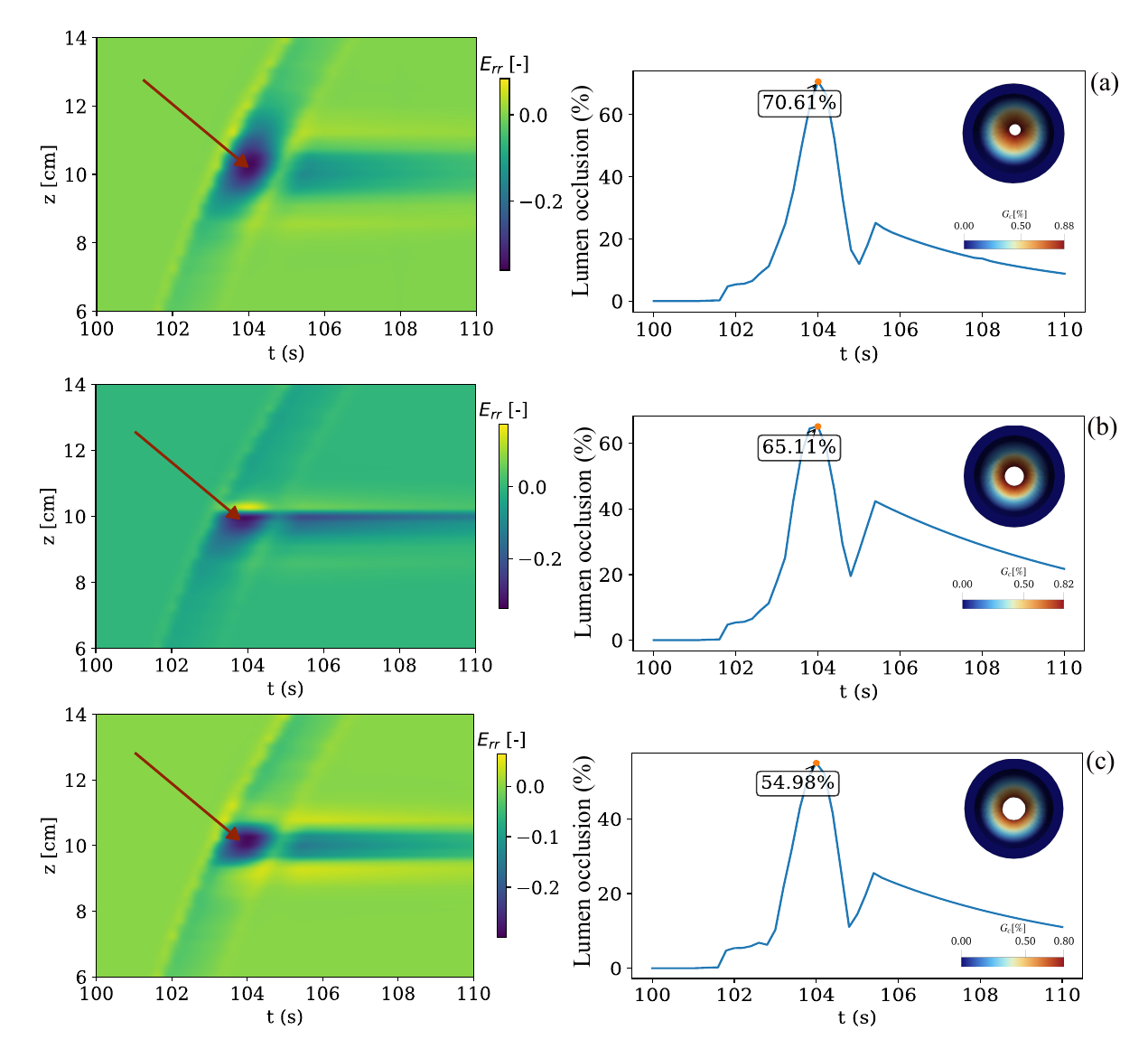}
    \caption{Comparison of transverse strain maps and lumen occlusion for different spike-burst activation profiles. (a) Gaussian, (b) hybrid Gaussian, and (c) Heaviside activation. The transverse strain maps represent the deformation perpendicular to the longitudinal axis and provide a numerical counterpart to the experimentally measured transverse maps \citep{kuruppu2024electromechanical}. Negative strain values indicate transverse compression associated with circumferential contraction.}
     \label{quantitative}
\end{figure}

{Figure~\ref{quantitative} shows radial strain maps with corresponding lumen closure for the three activation profiles. Strain levels are consistent with experimental measurements \(\sim 36 \pm 4\%\). In particular, we highlight a progressive reduction of contractility, $34\%$ (a), $32\%$ (b), $28\%$ (c), as the spike-burst activation profile changes from Gaussian to Heaviside. Morevoer, negative radial strain values are also observed reflecting the transverse compression induced by circumferential contraction. This contribution induces lumen closures ranging from \(\sim55\%\) to \(71\%\).
%The Gaussian profile produces the largest lumen closure, followed by the hybrid Gaussian--Heaviside profile and finally the pure Heaviside profile. 
Such a variability can be attributed to the spatial localization of the activation. The Gaussian profile exhibits a pronounced peak, resulting in a highly localized contraction concentrated within a relatively small region thus promoting effective lumen occlusion.
In contrast, the Heaviside profile distributes its maximum activation over a much wider region reducing the ability of the wall to generate a strong localized occlusion.
}

\clearpage
\newpage

\subsection{Influence of luminal folds}
Complete occlusion of the intestinal lumen cannot be achieved in a perfect cylindrically symmetric geometry \citep{kumar2015instabilities, zhao2023elastic} since contraction remains circumferentially uniform and the lumen tends to collapse symmetrically without reaching complete occlusion. However, in physiological conditions, the inner surface of the intestine is not smooth but contains folds that naturally break such a symmetry and create preferential contact regions and facilitate lumen occlusion.

\begin{figure}[h!]
    \centering
    \includegraphics[width=\linewidth]{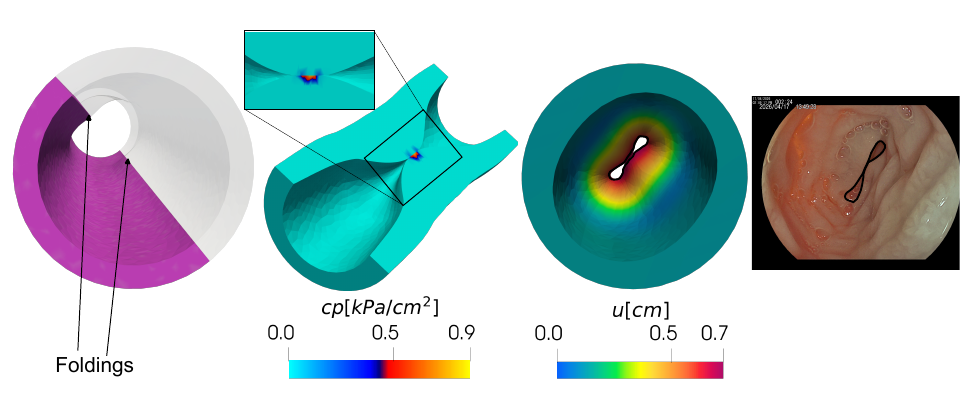}
    \caption{Role of luminal folds in intestinal occlusion. From left to right: internal view of the folded geometry generated from 2D Gaussian functions; contact pressure per unit surface ($cp$) and displacement magnitude ($u$) at maximum contraction with resulting figure-eight-shaped occlusion pattern (black contour) qualitatively consistent with the morphology observed in an endoscopic image of pig small intestine (right).}
    \label{full_closure_fold}
\end{figure}

To investigate this condition, we modified the reference geometry by introducing localized luminal folds generated from Gaussian functions. The geometry code was implemented in Fusion 360, and the resulting folded configuration is shown in Fig.~\ref{full_closure_fold}(left). Apart from this geometric modification, all model parameters and activation conditions were kept identical to those previously used for the myogenic case with Gaussian spike-burst activation.

The corresponding results are shown in Fig.~\ref{full_closure_fold}b showing the contact pressure $cp$ and the displacement magnitude $u$. As expected, the folded geometry successfully achieves complete lumen occlusion and implies complex self-contact and pressure maps. Noticeably, as the lumen collapses, the two folds progressively come into contact and generate a characteristic figure-eight-shaped occlusion pattern that  can be observed in the endoscopic {\it in vivo} pig intestine image in Fig.~\ref{full_closure_fold} (right). 

\begin{comment}
\begin{figure}[h!]
    %\centering
    \raggedright
    \hspace*{-2cm}
    \begin{subfigure}[t]{0.45\textwidth}
        \centering
        \includegraphics[height=5cm]{geom11.pdf}
        \caption{}
        \label{geom}
    \end{subfigure}
   % \hfill
    \begin{subfigure}[t]{0.5\textwidth}
        \centering
        \includegraphics[height=7.5cm]{geom22.pdf}
        \caption{}
        \label{geom2}
    \end{subfigure}

    \caption{Role of luminal folds in complete intestinal occlusion. (a) Internal view of the folded geometry generated from Gaussian-based folds (black contours). (b) Contact pressure ($cp$) and displacement magnitude ($u$) at maximum contraction. The folds break the cylindrical symmetry and promote complete lumen closure through self-contact. The resulting figure-eight-shaped occlusion pattern (black contour) is qualitatively consistent with the morphology observed in the endoscopic image.}
    \label{full_closure_fold}
\end{figure}
\end{comment}
These findings highlight the importance of geometric asymmetry in the mechanics of intestinal occlusion. While strong contractile activity, fiber-layer segregation, and self-contact are necessary ingredients for realistic intestinal motility, they are not sufficient to guarantee complete lumen occlusion in a perfectly symmetric geometry. Instead, complete occlusion emerges from the combined effects of muscle activation, contact mechanics, and geometric asymmetries associated with the physiological folded lumen.

\section{Conclusion}
\label{conclu}

Reproducing gastrointestinal motility remains an open problem in the literature:the complex electrophysiology and a layered microstructural organization of the intestinal wall challenges existing theoretical and computational modeling approaches. Simplified homogenized muscle fibers organization, in fact, introduces an artificial mechanical competition within the different layers, thereby limiting the achievable contraction amplitude and preventing realistic lumen occlusion, i.e., peristalsis.

The present contribution proposed a new electromechanical framework that explicitly accounts for the histological organization of the intestinal wall by assigning circumferential and longitudinal fibers to their respective anatomical layers. The model combines slow wave propagation, localized spike-burst activation, active strain mechanics, and a self-contact formulation based on an augmented Lagrangian approach thus improving our predictability power. Several spatial spike-burst activation profiles were further introduced to investigate how the localization and asymmetry of neural activation may influence intestinal motility and lumen occlusion.

We finally deploy a novel paradigm for GI motility, experimentally identifying two characteristic deformation regimes, namely peristalsis and wrinkling. Our analysis demonstrated that the transition between these regimes is primarily governed by the balance between the effective circumferential and longitudinal contractions. Large peristaltic deformations develop when the circumferential contractility remains dominant, whereas wrinkling emerges when this dominance decreases owing to either spatial desynchronization between slow wave and spike-burst activities or local asymmetries in the spike-burst activation profile. 
These findings provide a mechanistic explanation for the coexistence of different contraction patterns observed in vivo on small intestine segments.

%
%Quantitative comparisons with the experimental measurements reported by \citet{kuruppu2024electromechanical} demonstrate good agreement, with predicted transverse strains ranging from $28\%$ to $34\%$, compared with the reported experimental value of $36 \pm 4\%$. Moreover, the proposed layer-specific formulation increases the predicted lumen closure from approximately $30\%$ for homogenized fiber architectures to nearly $80\%$. This substantial improvement confirms that segregating the two muscle fiber families effectively reduces mechanical competition and significantly enhances contractile efficiency, producing contraction amplitudes approximately twice those reported in several previous electromechanical models.

%
We further demonstrate that large muscle activation, layer-specific fiber organization, and self-contact alone are not sufficient to achieve complete intestinal occlusion in a perfectly axisymmetric cylindrical geometry. By introducing physiological luminal folds that break circumferential symmetry, the model successfully reproduces--for the first time to the best of the authors' knowledge--complete lumen occlusion together with the characteristic figure-eight-shaped occlusion pattern observed in endoscopic images. These results highlight, for the first time, the critical role of geometric asymmetry in intestinal mechanics and calls for the need of realistic computational domains, based on clinical images, merged with advanced material modeling.

Overall, the study provides a physiologically grounded and mechanically robust framework for rethinking GI motility in computational studies. Beyond improving contraction amplitudes, it establishes a unified computational framework capable of reproducing peristalsis, wrinkling, and complete lumen occlusion, thereby providing a paradigm for investigating physio-pathological disorders.

\subsection*{Limitations and Perspectives}

The present study involves limitations related to both modeling assumptions and computational aspects. In particular, accurately describing the spatial and temporal variability of material parameters would require high-resolution imaging data and advanced material characterization. Given the challenges associated with acquiring such data, future work will focus on implementing a global sensitivity analysis to identify the parameters that most strongly influence the model predictions. 
Spike-burst activation modeling has been assumed phenomenological and localized in space. Generalized spatiotemporal electrophysiological model will enable a consistent representation of spike propagation towards patient-specific predictability. 
Although a simple inflation test was performed in the present work to approximate the prestressed reference configuration, it is worth noting that the computational framework developed in this study inherently supports solid-solid contact interactions. This capability opens promising perspectives for future experimental-based investigations, where a clinical catheter balloon (barostat) is expected  to be used to assess a rigorous prestress state, further enabling direct comparison between computational predictions and experimental measurements.

\section*{CRediT authorship contribution statement}

\textbf{René Thierry Djoumessi:} Writing -- review \& editing, Writing -- original draft, Validation, Software, Methodology, Investigation, Formal analysis, Conceptualization. \textbf{Christopher Miller:} Writing -- review \& editing, Writing -- original draft, Conceptualization. \textbf{Nipuni Nagahawatte:} Resources, Writing -- review \& editing. \textbf{Marco Paggi:} Writing -- review \& editing, Writing -- original draft, Methodology, Conceptualization, Supervision. \textbf{Leo K. Cheng:} Resources, Writing -- review \& editing, Supervision. \textbf{Alessio Gizzi:} Writing -- review \& editing, Writing -- original draft, Validation, Methodology, Investigation, Formal analysis, Conceptualization, Supervision.

\section*{Declaration of competing interest}

The authors declare that they have no known competing financial interests or personal relationships that could have appeared to influence the work reported in this paper.

\section*{Acknowledgments}

The authors acknowledge the support of 
ERC Consolidator Grant support from the European Union’s Horizon Europe research and innovation programme under grant agreement No. 101170592 - MiGEM; 
Italian National Group for Mathematical Physics (GNFM-INdAM);
Catalyst: Leaders funding provided by the New Zealand Ministry of Business, Innovation and Employment and administered by Royal Society Te Apārangi; 
Auckland Bioengineering Institute.

\section*{Data availability}
Data will be made available on request.

\newpage
\bibliographystyle{elsarticle-harv}
\bibliography{sample_jmps_abbreviated}

\clearpage
\newpage

\appendix
\section{Electro-physiological model}
\label{electrophy_mo}
 The SMC and ICC layers are denoted by the indices $s$ and $i$, respectively. The interaction between the transmembrane potentials $u_s, u_i$ and the local recovery variables $v_s, v_i$ is described by a coupled reaction--diffusion system formulated as two FitzHugh--Nagumo--type equations:
\begin{linenomath}
\begin{subequations}
\begin{align}
\frac{\partial u_s}{\partial t} &= f(u_s)+D_s\nabla^2 u_s-v_s + F_s(u_s,u_i) \quad \textrm{on} \quad \Omega_0 \times [0,T], 
\\
\frac{\partial v_s}{\partial t}  &= \epsilon_s[\lambda_s (u_s-\beta_s)-v_s] \quad \textrm{on} \quad \Omega_0 \times [0,T], \\
\frac{\partial u_i}{\partial t} &= g(u_i)+D_i\nabla^2 u_i-v_i + F_i(u_s,u_i) + I_{stim}^i \quad \textrm{on} \quad \Omega_0 \times [0,T], \label{eq:14c}\\
\frac{\partial v_i}{\partial t}  &= \epsilon_i(z)[\lambda_i (u_i-\beta_i)-v_i] \quad \textrm{on} \quad \Omega_0 \times [0,T],
\end{align}
\label{electrophysiology}
\end{subequations}
\end{linenomath}
where:
\begin{linenomath}
\begin{subequations}
\begin{align}
f(u_s) &=  k_su_s(u_s-a_s)(1-u_s) \,, \quad
& F_s(u_s,u_i) = \alpha_s D_{si}(u_s-u_i) \,, \\
g(u_i) &=  k_iu_i(u_i-a_i)(1-u_i) \,, \quad
& F_i(u_s,u_i) = \alpha_i D_{is}(u_s-u_i) \,.
\end{align}
\label{electrophysiologyfun}
\end{subequations}
\end{linenomath}
Here, $I_{stim}^i$ is the external currents applied to the ICC; $D_s, D_i$ are the diffusivities; $\lambda_s, \lambda_i$ are the coupling factors between the membrane potential and recovery variable; $D_{si}, D_{is}$ are the diffusivities of the gap junctions between the two cell species; $k_i, k_s, a_s, a_i,\alpha_s, \alpha_i$ are phenomenological model parameters.
The parameter $\epsilon_i(\xi)$ represents a space-dependent excitability function, decreasing with distance from the pylorus $\xi$. More information concerning the model parameters can be found in \citep{aliev2000simple, gizzi2010electrical, DJOUMESSI2024116989}.

\subsection{Weak form of the Electrophysiology}
The electrical fields are defined in the Sobolev space:
$$
(u_s,u_i,v_s,v_i) \in \mathcal{W}^4 := \big[L^{2}\!\big([0,T];\, H^{1}(\Omega_0)\big)\big]^4.
$$
associated test functions vanish on Dirichlet boundaries:
$$
(\delta u_s, \delta u_i, \delta v_s, \delta v_i) \in \mathcal{W}_0^4 := \big[L^{2}\!\big([0,T];\, H^{1}_0(\Omega_0)\big)\big]^4.
$$

Multiply the strong form Eq.~\eqref{electrophysiology} by the test functions $(\delta u_s, \delta u_i, \delta v_s, \delta v_i)$, integrate over $\Omega_0$, using the divergence theorem and the no–flux condition, the variational problem reads as to find $(u_s,u_i,v_s,v_i)\in\mathcal{W}^4$ such that:
\begin{linenomath}
\begin{subequations}
\begin{align}
\int_{\Omega_{0}}\frac{d u_s}{d t}\delta u_s+\int_{\Omega_{0}}D_s\nabla u_s\cdot\nabla\delta u_s&=\int_{\Omega_{0}}I_{ion}^s (u_s,v_s,u_i)\delta u_s \,,\\
\int_{\Omega_{0}}\frac{d v_s}{d t} v_s &=\int_{\Omega_{0}}R_{s}(u_s,v_s)\delta v_s \,, \\
\int_{\Omega_{0}}\frac{d u_i}{d t}\delta u_i+\int_{\Omega_{0}}D_i\nabla u_i\cdot\nabla\delta u_i&=\int_{\Omega_{0}}I_{ion}^i (u_s,v_i,u_i)\delta u_i \,,\\
\int_{\Omega_{0}}\frac{d v_i}{d t}\delta v_i &=\int_{\Omega_{0}}R_{i}(u_i,v_i^)\delta v_i \,,
\end{align}
\label{Weak_elect}
\end{subequations}
\end{linenomath}
where the source terms in Eq.~\eqref{Weak_elect} are given by:
\begin{linenomath}
\begin{subequations}
\begin{align}
I_{ion}^i (u_s,v_i,u_i) &=  g(u_i)-v_i + F_i(u_s,u_i) + I_{stim}^i \,,\\
R_{i} (u_i,v_i) &= \epsilon_i(z)[\lambda_i (u_i-\beta_i)-v_i] \,, \\
I_{ion}^s (u_s,v_s,u_i) &=  f(u_s)-v_i + F_s(u_s,u_i)  \,,\\
R_{s} (u_s,v_s) &= \epsilon_s[\lambda_i (u_s-\beta_s)-v_s] \,.
\end{align}
\label{source term}
\end{subequations}
\end{linenomath}
Using an implicit backward Euler time integration scheme to Eq.~\ref{Weak_elect}, the corresponding residuals are defined in Eq.~\ref{Weak_elect_discret}.

%\clearpage
%\newpage
\subsection{Time discretizations of the electrophysiology}
\begin{linenomath}
\begin{subequations}
\begin{align}
\mathcal{E}_{u_s}^{n+1}&:=\int_{\Omega_{0}}\frac{u_s^{n+1}-u_s^{n}}{\Delta t}\delta u_s+\int_{\Omega_{0}}D_s\nabla u_s^{n+1}\cdot\nabla\delta u_s-\int_{\Omega_{0}}I_{ion}^s (u_s^{n+1},v_s^{n+1},u_i^{n+1})\delta u_s \,,\\
\mathcal{E}_{v_s}^{n+1}&:=\int_{\Omega_{0}}\frac{v_s^{n+1}-v_s^{n}}{\Delta t}\delta v_s -\int_{\Omega_{0}}R_{s}(u_s^{n+1},v_s^{n+1})\delta v_s \,, \\
\mathcal{E}_{u_i}^{n+1}&:=\int_{\Omega_{0}}\frac{u_i^{n+1}-u_i^{n}}{\Delta t}\delta u_i+\int_{\Omega_{0}}D_i\nabla u_i^{n+1}\cdot\nabla\delta u_i-\int_{\Omega_{0}}I_{ion}^i (u_s^{n+1},v_i^{n+1},u_i^{n+1})\delta u_i \,,\\
\mathcal{E}_{v_i}^{n+1}&:=\int_{\Omega_{0}}\frac{v_i^{n+1}-v_i^{n}}{\Delta t}\delta v_i -\int_{\Omega_{0}}R_{i}(u_i^{n+1},v_i^{n+1})\delta v_i \,,
\end{align}
\label{Weak_elect_discret}
\end{subequations}
\end{linenomath}
In a compact monolithic algorithm scheme, the problem read as: find $u_s$, $u_i$, $v_s$ and $v_i$ such that
\begin{linenomath}
\begin{equation}
\label{monolithic_res}
    \mathcal{E}^{n+1}(u_s,u_i,v_s,v_i;\delta u_s, \delta u_i, \delta v_s, \delta v_i):= \mathcal{E}_{u_s}^{n+1}+ \mathcal{E}_{v_s}^{n+1}+ \mathcal{E}_{u_i}^{n+1}+ \mathcal{E}_{v_i}^{n+1}=0 \,,
\end{equation}
\end{linenomath}
for all test functions $\delta u_s, \delta u_i, \delta v_s, \delta v_i$.

\clearpage
\newpage
\section{Numerical implementation}
\label{numerical}
\subsection{Strong form}
The strong form of the problem is defined by a set of nonlinear, coupled partial differential equations that prescribe mechanical equilibrium in terms of the displacement field $\bf u$ and the pressure $p$. In the undeformed configuration $\Omega_0$ they read:
\begin{linenomath}
\begin{subequations}
\begin{align}
    \nabla \cdot \bP &= {\bf 0} \,,  
    \quad \textrm{on} \quad \Omega_0 \times [0,T] \label{eqn:line-1} 
    \\
    {\rm log}J + \frac{p}{\kappa} &=0 \,, \quad 
    \textrm{on} \quad \Omega_0 \times [0,T] \label{eqn:line-2} 
\end{align}
\label{Strong_form}
\end{subequations}
\end{linenomath}
The associated mixed boundary conditions concerning the normal displacement and traction are additionally given by:
\begin{linenomath}
\begin{subequations}
\begin{align}
    \bu \cdot \bn &=  0 \,, 
    \quad \text{on} \quad \Gamma_{d} \times [0,T]
    \label{eq:bc1}
    \\
    \bP \bn -p_0J\bF^{-T} \bn &= {\bf 0} \,,
    \quad \text{on} \quad \Gamma_{n} \times [0,T]
    \label{eq:bc2}
    \\
    \bP \bn -\bR_{\rm rob}J\bF^{-T} \bu &= {\bf 0} \,,
    \quad \text{on} \quad \Gamma_{r} \times [0,T]
    \label{eq:bc3}
\end{align}
\label{BC}
\end{subequations}
\end{linenomath}
being $\Gamma_{D}$ and $\Gamma_{n}$ a disjoint partition of the boundary (see Fig. \ref{problem_config}): 
condition \eqref{eq:bc1} constraints normal motion along $\Gamma_{D}$;
the term $p_0$ in \eqref{eq:bc2} denotes a prescribed (possibly time-dependent) boundary load (normal stress--pressure) associated with the presence of digesta within the lumen;
the term $\bR_{\rm rob}$ in \eqref{eq:bc3} prescribes the stiffness associated with the presence of the surrounding organs and mesentery.   

\subsection{Weak formulation of the problem}

The trial spaces for the displacement and pressure fields, in which the weak formulation of the problem is defined, are given by:
\begin{linenomath}
\begin{equation}
\bu, \delta \bu \in \mathbf{V}:=L^2(0,T;\mathbf{H}^1_{\Gamma_{d}}(\Omega_0)), \; p, \delta p \in Q:= L^2(0,T; L^2(\Omega_0))
\label{eq:14}
\end{equation}
\end{linenomath}

Accordingly, Eq.~\eqref{Strong_form} with associated boundary conditions \eqref{BC}, define the weak formulation of the problem allowing the calculation of displacement $\bu$ and pressure and $p$ such that
\begin{equation}
\begin{aligned}
\mathcal{M}(\bu,p;\delta \bu,\delta p) :=
&\int_{\Omega_{0}}\bP:\nabla\delta \bu
-\int_{\Gamma_{N}} p_{0}(t)J\bF^{-T} \bn \cdot \delta \bu
\\
&+\int_{\Omega_{0}}\left(\log J + \frac{p}{\kappa}\right)\delta p 
+\int_{\Omega_{0}}\zeta_{stab}\nabla p \cdot \nabla \delta p
=0 \,,
\end{aligned}
\label{weakform}
\end{equation}
where $\zeta_{stab}$ is a positive pressure stabilization parameter to fufill the \textit{inf-sup} stability condition~\citep{propp2020orthotropic} since the unknowns were approximated using Lagrangian shape functions, employing $\mathcal{P}_2$ elements for the displacement field and $\mathcal{P}_1$ elements for the pressure field.

\subsubsection{Contact Model}
\label{sec:contactAlg}
The first-order optimality conditions of the augmented formulation of the contact model \eqref{lag_f} yields the following weak form:
\begin{equation}
\begin{aligned}
%\linenomath
& \mathcal{D}\mathbb{P}(\bu, p)[\delta \bu] 
- \int_{\Gamma_c^S} \lambda_\bN \, \bn_\bx \cdot \left( \delta \bu(\bX) - \delta \bu(\bY) \right) \, d\Gamma = 0 
\quad \forall \, \delta \bu, \\
%\linenomath
& - \frac{1}{r_p} \int_{\Gamma_c^S} \left( \lambda_\bN + \left[ \lambda_\bN + r_p g \right]_{-} \right) 
\delta \lambda_\bN \, d\Gamma = 0 
\quad \forall \, \delta \lambda_\bN, \\
& \mathcal{D}\mathbb{P}(\bu, p)[\delta p]=0 \quad \forall \, \delta p\, \
\label{lag_contact}
\end{aligned}
\end{equation}
with $\mathcal{D}\mathbb{P}(\bu,p)[\delta \bu]$ and $\mathcal{D}\mathbb{P}(\bu,p)[\delta p]$ the directional derivative of the potential energy with respect to a virtual displacement $\delta \bu$ and the virtual pressure $\delta p$ respectively while $\delta \lambda_\bN$ is a virtual variation of the contact multiplier field.

The chosen formulation owns several advantages:
(1) it avoids explicit inequality constraints on $\lambda_\bN$;
(2) it remains differentiable ``almost'' everywhere, which enables the use of a generalized Newton method;
(3) it involves only the surface normal $\bn_\bx$, which is differentiable and straightforward to compute numerically;
(4) the method is unbiased and does not require differentiating between master and slave surfaces when searching for potential contact points.

Accordingly, the whole weak contact mechanical problem \eqref{weakform} generalizes as:

\begin{equation}
\begin{aligned}
\mathcal{M}(\bu,\bu_b, p, \lambda_\bN;\delta \bu, \delta \bu_b, \delta p, \delta \lambda_\bN) &:= \int_{\Omega_{0}}\bP:\nabla\delta \bu \; \mathrm{d} V   -\int_{\Gamma_{N}} p_{0}(t)J\bF^{-T} \bN \cdot \delta \bu \; \rm d \Gamma \\
    &- \int_{\Gamma_c^S} \lambda_\bN \, \bn_\bx \cdot \left( \delta \bu(\bX) - \delta \bu(\bY) \right) \; \rm d \Gamma\\
    &- \frac{1}{r_p} \int_{\Gamma_c^S} \left( \lambda_\bN + \left[ \lambda_\bN + r g \right]_{-} \right) 
     \delta \lambda_\bN \, \rm d\Gamma\\
    &+\int_{\Omega_{0}}\left({\rm log}J + \frac{p}{\kappa}\right)\delta p  \; \rm d V
    + \int_{\Omega_{0}}\zeta_{stab}\nabla p \cdot \nabla \delta p \; \mathrm{d} V\\
    &+\int_{\Gamma_r} J\, \bR_{\rm rob} \cdot \bF^{-T} \bu \cdot \delta \bu \; \rm d \Gamma = 0 \,.
\end{aligned}
\label{mech_problem}
\end{equation}

\subsection{Finite element and time discretisations}
The numerical implementation was carried out using the finite element open-source software \texttt{GetFEM}, which offers a flexible model-based architecture well suited for multiphysics problems. A staggered algorithm was employed to solve the coupled problem by defining two distinct models in on model, corresponding to the electrophysiological problem (see the Residual in \eqref{monolithic_res}; and the second, representing the mechanical problem. Although a staggered approach was used to couple the electrophysiology and mechanical problems, each subproblem was solved using a monolithic algorithm.

%This modular strategy provides significant flexibility, notably allowing the mesh used for the electrophysiological problem to differ from the one used for the mechanical problem. While this capability is not the focus of the present article, it constitutes an important feature of the proposed computational framework. Another key advantage lies in the ability of the framework to define contact variables on specific surfaces through filtered variables. These variables are associated with a finite element space but restricted to a prescribed region, such that only the corresponding degrees of freedom are active. This provides additional flexibility, allowing the choice of finite element order and integration scheme independently from the rest of the problem. This feature was used to define the Lagrange multiplier $\lambda_N$ in the contact formulation and was defined with $\mathcal{P}_4$ finite element.  

The nonlinear mechanical problem is solved using the Newton-Raphson method coupled with a line search algorithm to ensure convergence and, at each Newton's iteration, the resulting linear system: 
$$d \mathcal{M}(\Delta \bu, \Delta p,\Delta \lambda_\bN;  \delta \bu,\delta p, \delta \lambda_\bN)=-\mathcal{M}(\bu_k^{n+1},p_k^{n+1}, \lambda_{\bN_k}^{n+1};\delta \bu,\delta p, \delta \lambda_\bN)$$ 
is solved for the corrections $\alpha\Delta \bu$,  $\alpha \Delta p$, $\alpha \Delta \lambda_\bN$ using a parallel \texttt{MUMPS} solver. The parameter $\alpha$ is computed using the simplest line search algorithm. The electrophysiological problem was solved via the \texttt{GMRES} solver with the preconditioner \texttt{ILU} factorization.

To avoid numerical difficulties associated with the discontinuous derivative of the Heaviside function in the Newton solver, the parameters were regularized  using a first-degree polynomial interpolation within a narrow transition region in the thickness, ensuring differentiability and improving convergence of the nonlinear solver.
The algorithm for the solution of the coupled electromechanical problem describing the intestinal motility with contact is detailed in \ref{algoAp}.

\clearpage
\newpage
\section{Material parameters and thickness distribution}
\label{param}
\begin{table}[h!]
    \centering
    \caption{Material parameters of the active strain model \citep{brandstaeter2018computational}.}
    \begin{tabular}{c c c c c c}
    \hline
     $\alpha_c$  & $\alpha_l$ & $\beta_1$ & $\beta_2$ & $u_s^{th}$ & $u_{s,p}^{th}$\\
     \hline
        $0.5$ & $0.1$ & $10$ & $10$ & $25\% u_s$ & $2\% u_{s,p}$\\
    \hline
    \end{tabular}
    \label{tab:active}
\end{table}
\begin{table}[h!]
\centering
\caption{Material parameters used for the uni-axial test \citep{nagaraja2021phase}}
\begin{tabular}{c c c c c c c c} 
 \hline
$\mu [\rm kPa]$ & $k_1^l[\rm kPa]$ & $k_2^l[-]$ & $k_1^c[\rm kPa]$ & $k_2^c[-]$ & $k_1^d[\rm kPa]$ & $k_2^d[-]$ & $\theta [\circ]$\\
 \hline
  $1.5$ & $3.13$ & $1.18$ & $5.78$ & $0.0199$ & $3.65$ & $0.31$ & $38.18$\\
 \hline
\end{tabular}
\label{table:mechanical}
\end{table}
%
%\section{Distribution of material parameters across the thickness.}
%\label{sec:A}
\begin{figure}[ht!] 
\centering
\includegraphics[width=\textwidth]{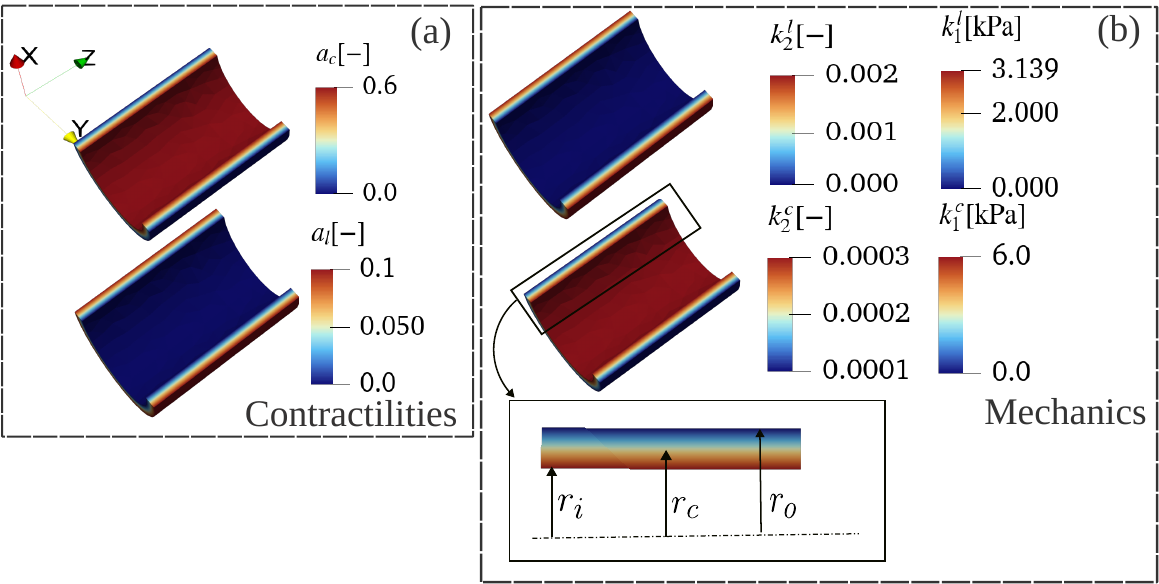}
\caption{Generic representation of the through-thickness parameter distributions
following Eq.~\eqref{func_c} with $r_c = 1.35$:
(a) contractility parameter distribution;
(b) anisotropic mechanical parameter distribution.}
\label{material_distru}
\end{figure}

\clearpage
\newpage
\section{Shape of the spike burst used in the paper}
\label{sec:B}

The spatial activation profile of the Gaussian and the hybrid Gaussian-Heaviside profile is then given in the following equations:
$M_z(\bx)=\exp\!\left(-{(z-z_0)^2}/{2\sigma_z^2}\right)$ and $M_z(\bx)=\exp\!\left(-\frac{(z-z_0)^2}{2\sigma_z^2}\right)\,H(z_c - z)$. In  which the parameter $z_c$ controlling the transition between the two regions. 

\begin{figure}[ht!] 
\centering
\includegraphics[width=\textwidth]{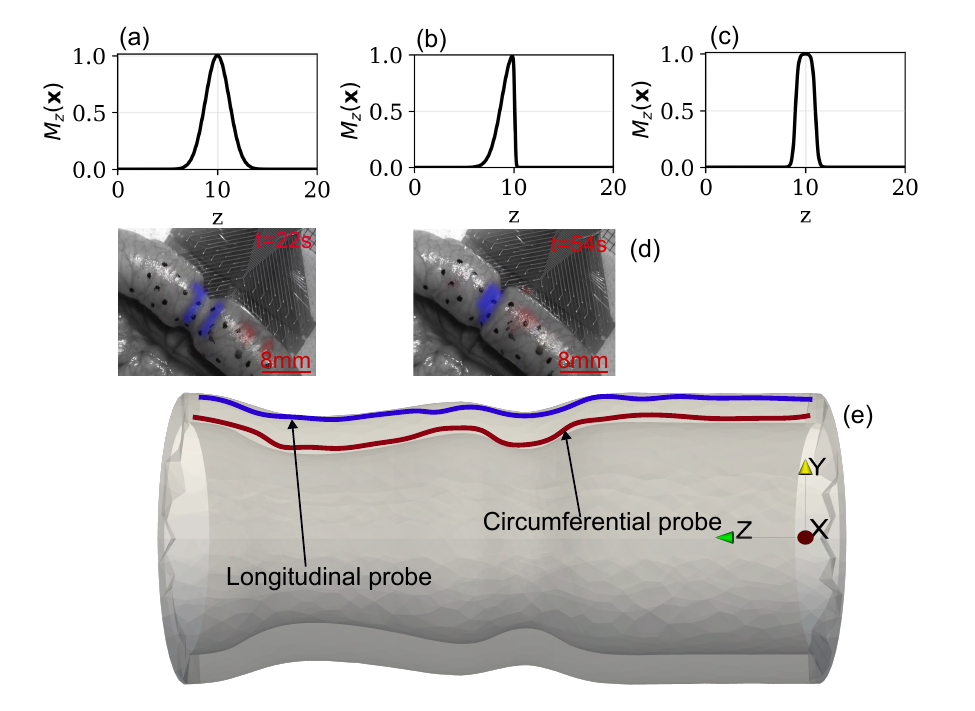}
\caption{Shapes of the functions used in this study to impose spike burst excitation: (a) Gaussian profile, (b) Hybrid Gaussian–Heaviside profile, and (c) Heaviside profile. (d) Representative frames extracted from the experimental recordings of \citet{kuruppu2024electromechanical} illustrating the spatial variability of spike-burst-induced contractions. The blue regions indicate the location of the spike-burst activity, while the corresponding deformation of the intestinal wall is visible in the surrounding tissue,  (e) probe use to extract the quantitative data in circumferential and longitudinal layer.}
\label{maks}
\end{figure}

\clearpage
\newpage
\section{Numerical algorithm}
\label{algoAp}

\begin{algorithm}
\caption{Electromechanical motility algorithm for solving Eq.~\ref{mech_problem} and Eq.~\ref{monolithic_res}}
\label{alg:cap}
\begin{algorithmic}[1]

\State \textbf{Input:} Initial and boundary conditions.

\While{$t^n < T$}
    \State \textbf{Given} $(\bu^n,p^n,\lambda_\bN^n)$, solve:
    \[
    \mathcal{E}(u^{n+1}_s,u^{n+1}_i,v^{n+1}_s,v^{n+1}_i;\delta u_s,\delta u_i,\delta v_s,\delta v_i)=0
    \]
    \State Update $(u_s^n,u_i^n,v_s^n,v_i^n)\gets(u_s^{n+1},u_i^{n+1},v_s^{n+1},v_i^{n+1})$

    \State \textbf{Solve mechanical problem (Newton):}
    \For{$k$}
        \State Solve:
        \[
        d\mathcal{M}(\Delta\bu,\Delta p,\Delta\lambda_\bN;\delta\bu,\delta p,\delta\lambda_\bN)
        =-\mathcal{M}(\bu_k^{n+1},p_k^{n+1},\lambda_{\bN,k}^{n+1};\delta\bu,\delta p,\delta\lambda_\bN)
        \]
        \State $\alpha\gets1$
        \While{criterion not satisfied}
            \[
            \bu^\text{trial}=\bu_k^{n+1}+\alpha\Delta\bu,\quad
            p^\text{trial}=p_k^{n+1}+\alpha\Delta p,\quad
            \lambda_\bN^\text{trial}=\lambda_{\bN,k}^{n+1}+\alpha\Delta\lambda_\bN
            \]
            \State Check $\mathcal{M}(\bu^\text{trial},p^\text{trial},\lambda_\bN^\text{trial})$
            \State Reduce $\alpha$ if needed
        \EndWhile
        \State Update:
        \[
        \bu_{k+1}^{n+1}=\bu_k^{n+1}+\alpha\Delta\bu,\quad
        p_{k+1}^{n+1}=p_k^{n+1}+\alpha\Delta p,\quad
        \lambda_{\bN,k+1}^{n+1}=\lambda_{\bN,k}^{n+1}+\alpha\Delta\lambda_\bN
        \]
        \If{$\|\bu_{k+1}^{n+1}-\bu_k^{n+1}\|<tol$}
            \State break
        \Else
            \State $k\gets k+1$
        \EndIf
    \EndFor

    \State $(\bu^{n+1},p^{n+1},\lambda_\bN^{n+1})\gets(\bu_{k+1}^{n+1},p_{k+1}^{n+1},\lambda_{\bN,k+1}^{n+1})$
    \State $t\gets t+\Delta t$
    \State \textbf{Output:} $\bu^{n+1},p^{n+1}$ and EP variables

\EndWhile
\end{algorithmic}
\end{algorithm}
\clearpage
\newpage
\begin{comment}
\section{Result for independent spike bursts}

\clearpage
\newpage
\section{Results For synchronized spikes}
\label{synchospike}
\begin{figure}[h]
\centering\includegraphics[width=\textwidth]{Result__synch_new.pdf}
    \caption{Temporal and spatial evolution of the slow wave $u_s$, the spike burst $u_{s,p}$, the displacement $u$, and the effective $G_c$. The results show a strong localized peristaltic contraction and wrinkling of the intestinal wall. They also reveal a synchronized wave propagation between $u_s$ and $u_{s,p}$. The parameters used in this simulation are $z_0 = 10\,\mathrm{cm}$, $\sigma_z = 1.2$, $\tau_{\mathrm{on}} = 8$, $A_p = 0.45$, and $u_{\mathrm{tr}} = 0.2$.}
    \label{synchrospikeburst}
\end{figure}
\end{comment}

%\section{Result of the Gaussian function}
%\label{gauss}
%\begin{figure}[h]
%\centering\includegraphics[width=0.8\textwidth]{images/results_gauss.pdf}
%    \caption{Solution showing the two regimes. PS: peristalsis and  WK: wrinkling regime.} 
%     \label{result_gauss}
%\end{figure}

%\section{Result of the Gaussian-step function}
%\label{gaussstep}
%\begin{figure}[h]
%\centering\includegraphics[width=0.8\textwidth]{images/result_brute1.pdf}
%    \caption{Solution showing the two regimes. PsW: cohesistance of peristalsis and Wrinkling, WK: wrinkling regime.} 
%     \label{result_br}
%\end{figure}

\newpage
\clearpage
\section{Result of the Heaviside smooth function}
\label{heavisidefunction}

\begin{figure}[h!]
\centering\includegraphics[width=\textwidth]{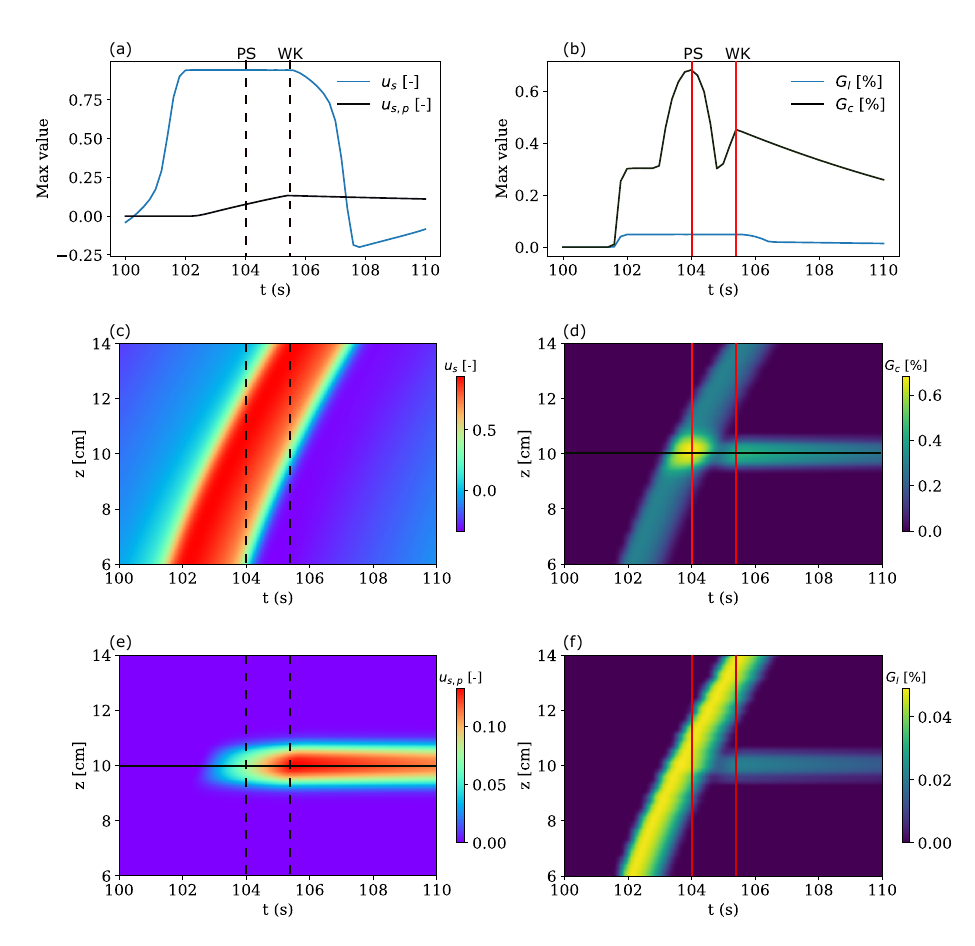}
    \caption{Quantitative analysis of the interaction between slow waves, spike-burst activity, and intestinal muscle contraction. (a) Temporal evolution of the maximum amplitudes of the slow wave $u_s$ and spike-burst $u_{s,p}$ activity. (b) Maximum effective contractility of the longitudinal $G_l$ and circumferential $G_c$ layer.  (c-f) Spatio-temporal distributions of $u_s$, $G_c$, $u_{s,p}$, and $G_l$ extracted along the selected probes (see \ref{sec:B}). Vertical lines identify two distinct deformation regimes: the PS regime, corresponding to the coexistence of peristaltic contraction and the WK regime, characterized by wrinkling-dominated behavior. Horizontal lines identify the symmetric of $u_{s,p}$ and  $G_c$ in space}.
     \label{comparative_heaviside}
\end{figure}

%\clearpage
%\newpage

%\section{Experimental data}
%\label{experi}

%\begin{figure}[h]
%    \centering
%    \includegraphics[width=0.9\linewidth]{images/experiment.pdf}
%    \caption{Experimental measurement of the strain from %\citep{kuruppu2021high}}
%    \label{experiemnet}.
%\end{figure}

\end{document}